\pdfoutput=1
\documentclass[aps,prd,amsmath,floats,floatfix, twocolumn,
superscriptaddress,nofootinbib,showpacs, amssymb,
]{revtex4-2}

\usepackage[T1]{fontenc}
\usepackage[utf8]{inputenc}
\usepackage{lmodern}
\usepackage{subfig}
\usepackage{xfrac}
\usepackage{verbatim}

\usepackage[dvipsnames, usenames]{xcolor}
\definecolor{linkcolor}{rgb}{0.0,0.3,0.5}
\usepackage[hypertexnames=false, unicode, colorlinks=true, linkcolor=linkcolor,
citecolor=linkcolor, filecolor=linkcolor,urlcolor=linkcolor,
pdfusetitle]{hyperref}

\usepackage[all]{hypcap}
\usepackage{graphicx}
\usepackage{xspace}
\usepackage[normalem]{ulem} %for \sout
\usepackage{bm} % boldmath
\usepackage{enumitem}
\usepackage{eqnarray}
\usepackage{caption}
\usepackage{microtype}

\usepackage[english]{babel}

\usepackage{blindtext}
\usepackage{orcidlink}

\graphicspath{%
  {figs/}%
}

\DeclareMathAlphabet{\mathpzc}{OT1}{pzc}{m}{it}

\newlist{todolist}{itemize}{2}
\setlist[todolist]{label=$\square$}

\begin{document}

\title{Statistical identification of ringdown modes with rational filters}

\newcommand\Caltech{\affiliation{TAPIR, California Institute of Technology, Pasadena, CA 91125, USA}}
\newcommand{\Perimeter}{\affiliation{Perimeter Institute for Theoretical Physics, Waterloo, ON N2L2Y5, Canada}}
\newcommand\anu{\affiliation{OzGrav-ANU, Centre for Gravitational Astrophysics, College of Science, The Australian National University, ACT 2601, Australia}}

\author{Neil Lu \orcidlink{0000-0002-8861-9902}}
\email{neil.lu@anu.edu.au}
\anu

\author{Sizheng Ma\orcidlink{0000-0002-4645-453X}}
\email{sma2@perimeterinstitute.ca}
\Perimeter

\author{Ornella J. Piccinni\orcidlink{0000-0001-5478-3950}}
\email{ornellajuliana.piccinni@anu.edu.au}
\anu

\author{Ling Sun\orcidlink{0000-0001-7959-892X}}
\email{ling.sun@anu.edu.au}
\anu

\author{Eliot Finch\orcidlink{0000-0002-1993-4263}}
\Caltech

% Because hyperref only gets the *last* author, we need to be explicit.
\hypersetup{pdfauthor={Lu et al.}}

\date{\today}

%==========================================================================
\begin{abstract}
Measuring quasinormal modes (QNMs) during the ringdown phase of binary black hole coalescences provides key insights into merger dynamics and enables tests of the no-hair theorem. The QNM rational filter has recently been introduced as a technique to identify specific QNMs in ringdown signals without sampling over mode amplitudes and phases. In this work, we extend the QNM rational filter framework to quantify the statistical confidence of subdominant mode detections in real gravitational wave (GW) observations. We employ a frequentist approach to estimate false-alarm probabilities and propose a workflow for robust identification of specific QNMs. We first validate our methodology using synthetic signals generated from numerical relativity waveforms. We then reanalyze the first GW event, GW150914, finding a marginal detection of an overtone, but at time when the applicability of constant amplitude QNM fits is not fully understood. This extended methodology provides a systematic approach to improving the reliability of QNM detections, paving the way for more precise tests of strong-field gravity with current and future GW observations.
\end{abstract}

\maketitle

%\done
%==========================================================================
\section{Introduction}
\label{sec:introduction}
The ringdown phase of a binary black hole (BH) coalescence corresponds to the relaxation of the deformed remnant BH as it settles into an equilibrium state. At sufficiently late times, the dynamics are expected to be linear, with the emitted gravitational waves (GWs) well described by a superposition of quasinormal modes (QNMs), each contributing as a damped sinusoid to the signal \cite{vishveshwara1970, teukolsky1973, chandrasekhar1975, leaver1985, kokkotas1999,berti2009}. Due to the no-hair theorem in general relativity (GR)~\cite{israel1967, carter1971, hawking1972, robinson1975, mazur1982}, the complex frequencies of these QNMs are uniquely determined by the mass and spin of the remnant BH. Independently measuring multiple QNMs and inferring the corresponding mass and spin --- a process known as BH spectroscopy --- provides a means to test the no-hair theorem \cite{dreyer2004, berti2006, isi2019} and the BH area law \cite{isi2021}. This is made possible by recent detections of binary BH coalescences with ground-based detectors, like LIGO \cite{collaboration2015}, Virgo \cite{acernese2014}, and KAGRA \cite{akutsu2020}, which allow for the possibility to measure multiple QNMs from individual events. 

At present, various methods are used to extract QNMs from observational data, e.g., time-domain \cite{carullo2019, isi2019, isi2021a, ma2022, siegel2025, dong2025}, a wavelet-based frequency-domain \cite{finch2021,finch2022}, using ``gating and in-painting'' \cite{capano2023,wang2023}, parameterized effective one-body \cite{brito2018, maggio2023, mihaylov2023, vandemeent2023} and machine-learning based techniques \cite{pacilio2024a}.
% \red{\sout{The diversity of fitting techniques exists because matched filtering, which operates in the frequency domain, cannot be implemented for signals with acyclic boundary conditions. These acyclic boundary conditions occur because it is necessary to exclude the earlier part of the waveform, the merger phase, from the analysis. Different ringdown analysis methods use different approaches to truncate the data and handle the high-frequency artifacts which may arise. The starting time of the ringdown phase is also uncertain, further complicating the analysis}} 
Different methods are used to avoid spectral leakage while processing short ringdown signals under acyclic boundary conditions. They have been applied to GW events such as GW150914 \cite{ligoscientificcollaborationandvirgocollaboration2016, ligoscientificandvirgocollaborations2016, carullo2019,bustillo2021,ghosh2021,finch2022, ma2023,ma2023a,crisostomi2023,wang2023,wang2024,correia2024}, GW190521 \cite{ligoscientificcollaborationandvirgocollaboration2020, capano2023, siegel2023, capano2024}, and a selected catalog of GW events \cite{ghosh2021, theligoscientificcollaboration2021}. While individual pipelines claim the detection of an overtone for GW150914 \cite{isi2019} or a higher-order mode for GW190521 \cite{capano2023,siegel2023, capano2024}, there is a lack of universal consensus about these detections, see particularly \cite{cotesta2022,carullo2023,isi2023, siegel2023}. 

The QNM rational filter has lately been proposed as a new method to identify QNMs through mode cleaning \cite{ma2022,ma2023,ma2023a}. The filter removes particular QNM(s) from a ringdown signal and depends only on the mass and spin of the remnant BH (by assuming that GR is correct). Crucially, the filter does not sample over the QNM amplitude(s) and phase(s). 
%Under certain prior assumptions~\cite{ma2023a}, the filter is related to the time-domain analysis in \cite{carullo2019, isi2019, isi2021a} with maximum likelihood estimation for the mode amplitudes. 
Previous studies with QNM rational filters have developed a hybrid Bayesian-like framework by formulating a novel likelihood function and have tentatively applied the framework to GW150914 in a search for the first overtone \cite{ma2023a, ma2023}.
The similarities and differences between the QNM rational filter and other time-domain analyses are discussed in Appendix~\ref{sec:semi-Bayesian} and the Supplemental Material of Ref.~\cite{ma2023a}. 
However, the statistical properties and robustness of the filter in the low signal-to-noise ratio (SNR) regime remain unexplored.

In this paper, we extend our previous studies \cite{ma2022,ma2023,ma2023a} to quantify the statistical significance of QNM observations made with the rational filter. In particular, we focus on the detection of subdominant ringdown modes with low SNRs. We take a frequentist approach to estimate false-alarm probabilities (FAPs) caused by background noise. Building on this, we formulate a workflow to robustly identify QNMs in ringdown signals and apply the workflow to simulations using numerical-relativity (NR) waveforms as well as the real event, GW150914. 

The structure of the paper is organized as follows. In Sec.~\ref{sec:background}, we review the background of QNMs, the rational filter, and define a detection statistic. In Sec.~\ref{sec:FAP}, we quantify the significance of the observation of a QNM via a FAP and describe ways to incorporate information obtained from inspiral-merger-ringdown (IMR) analysis that can increase the sensitivity to subdominant modes. In Sec.~\ref{sec:SNR}, we demonstrate a procedure of estimating ringdown SNR using a relationship between the detection statistic and SNR. In Sec.~\ref{sec:analysis_procedure}, we formulate a full workflow for analyzing GW events to identify the QNM contents. In Sec.~\ref{sec:NR}, we apply the workflow to NR simulations and demonstrate the expected modes can be successfully identified following such a procedure. We reanalyze GW150914 in Sec.~\ref{sec:GW150914} to quantify the statistical significance of the first overtone. Finally, we conclude and discuss future work in Sec.~\ref{sec:conclusion}.

%==========================================================================
\section{QNM rational filter}
\label{sec:background}
% The ringdown signal consists of GW harmonics $h_{lmn}(t)$. Within the linear regime, their time evolution is given by:
The real-valued time-series of the ($\ell,  m, n$) QNM in a GW detector is given by:
\begin{gather}
    h(t) = A_{\ell mn} e^{-(t - t_0)/\tau_{\ell mn}} \times \cos \left[ 2\pi f_{\ell mn} (t-t_0) + \phi_{\ell mn} \right] \,, \label{eq:QNM_def}
\end{gather}
% The time evolution of the $(\ell, m, n)$ QNM is given by $A_{\ell mn} e^{-i[\omega_{\ell mn}(t-t_0) + \phi_{\ell mn}]}$,
where $t_0$ is a given reference time, 
% The complex QNM frequency is given by $\omega_{\ell mn} = 2\pi f_{\ell mn} - {i}/{\tau_{\ell mn}}$ is characterized by an 
$f_{\ell mn}$ is the oscillation frequency and $\tau_{\ell mn}$ is the damping time. In GR, ($f_{\ell mn}, \tau_{\ell mn}$) are uniquely determined by the mass and spin of the remnant Kerr BH, whereas the amplitude $A_{\ell mn}$ and phase $\phi_{\ell mn}$ encode information about the merger dynamics. 

Throughout this work, we use three indices $(\ell,m,n)$ to label QNMs\footnote{QNMs are typically labeled by four indices; see the Appendix of \cite{giesler2024} for a brief summary.}. 
We adopt the convention $f_{\ell mn}>0$, with $m>0$ ($m<0$) indicating the prograde (retrograde) modes~\cite{Berti:2025hly,giesler2024,li2022}.
In Eq.~\eqref{eq:QNM_def} and throughout this paper, we consider only the prograde modes\footnote{The strain in Eq.~\eqref{eq:QNM_def} contains two prograde modes, emitting towards both North and South directions with respect to the BH spin axis.} and neglect retrograde modes.
%, which implies $m>0$ (\red{since we use convention 2 as described in \cite{giesler2024}}, see \cite{li2022} \red{for additionally clarification} about notations). 
To avoid word clutter, we refer to modes by their mode numbers, i.e. the 220 mode refers to the ($\ell=2$, $m=2$, $n=0$) mode.
We denote time $t$ in units of the final remnant BH mass $M_{\rm rem}$. 
% \red{\sout{In this case, we assume $m>0$ and use the labeling convention $h_{lmn}$ ($h_{l-mn}$) to represent the harmonic components emitted towards the north (south) with respect to the system.}}
% In GR, the complex frequencies $\omega_{\ell mn}$ depend only on the mass and spin of the remnant BH. In contrast, the amplitudes $A_{lmn}$ and phases $\phi_{lmn}$ depend on the initial conditions of the binary BH inspiral, which affect the perturbation of the remnant. 
%In GW observations, a QNM mode will consist of both complex harmonics $\pm m$ and we will label the ($\ell$, $m$, $n$) QNM to refer to both harmonics $\pm m$. 
%\sma{You are missing angular dependences}
%\begin{gather}
%    h(t) = \sum_{lmn} h_{l m n}(t) \, . \label{eq:QNM_superposition}
%\end{gather}

The QNM rational filter removes the QNM in Eq.~\eqref{eq:QNM_def} by multiplying the signal with the frequency-domain filter \cite{ma2022, ma2023, ma2023a}:
\begin{eqnarray}    
    \mathcal{F}_{\ell mn}(\omega; M_f, \chi_f) = \frac{\omega - \omega_{\ell mn}}{\omega - \omega^*_{\ell mn}} \, \frac{\omega + \omega^*_{\ell mn}}{\omega + \omega_{\ell mn}} \, ,\label{eq:filter}  %\nonumber 
%    = \frac{f-f_{lmn} + \frac{i}{2\pi \tau_{lmn}}}{f-f_{lmn} - \frac{i}{2\pi \tau_{lmn}}} \, \frac{f+f_{lmn} + \frac{i}{2\pi \tau_{lmn}}}{f+f_{lmn} - \frac{i}{2\pi \tau_{lmn}}} \, , \label{eq:filter}
\end{eqnarray}
where $\omega$ is the real-valued frequency, $\omega_{\ell mn} = 2\pi f_{\ell mn} - {i}/{\tau_{\ell mn}}$ is the complex QNM frequency, $M_f$ and $\chi_f$, are a given remnant BH mass and spin, and $^*$ represents the complex conjugate.
% $\omega_{\ell mn}$ refers to a specific QNM frequency which depends on the final remnant BH mass $M_{\rm rem}$ and spin $\chi_{\rm rem}$, 
A linear superposition of multiple QNMs can be simultaneously removed using a total filter:
\begin{equation}
\mathcal{F}_\text{tot}(\omega; M_f, \chi_f) = \prod_{\ell mn} \mathcal{F}_{\ell mn}(\omega; M_f, \chi_f) \, . \label{eq:total_filter}
\end{equation}

Because the filter is applied to the data in the frequency domain, we transform the GW time-series data, $d$, into the frequency domain (denoted by $\tilde{d}$) using a discrete Fourier transform:
\begin{equation}
    \tilde{d}_k = \frac{1}{\sqrt{N}} \sum_{n=0}^{N-1} d_{{n}} e^{-i 2\pi  \frac{k}{N}n} \, , \label{eq:fft}
\end{equation}
where $n$ is the index of the time-series, $N$ is the total number of data points, and we use the orthogonal normalization convention. After applying the filter in the frequency domain, we convert the filtered data back to the time domain using an inverse Fourier transform:
\begin{equation}
d^F_n = \frac{1}{\sqrt{N}} \sum_{k=0}^{N-1} \mathcal{F}_{\ell mn}\left(2\pi \frac{k}{N}\right) \tilde{d}_k e^{i2\pi \frac{k}{N}n} \, , \label{eq:filtered_data}
\end{equation}
where the superscript ``$F$'' denotes the filtered data.
The rational filter completely removes the mode(s) from the time-series if the correct BH parameters are used to construct the filter. 

In Refs.~\cite{ma2023, ma2023a}, analysis using the rational filter uses a likelihood function:
\begin{gather}
    \text{ln } \mathcal{L}(d_t | M_f, \chi_f, \mathcal{F}_\text{tot}, \Delta t_0) = -\frac{1}{2} \sum_{i,j>t_0} d_i^F C_{ij}^{-1} d_j^F \, , \label{eq:likelihood}
\end{gather}
where $\Delta t_0=t_\text{analysis} - t_0$ with $t_\text{analysis}$ as the starting time of the analysis and we take $t_0$ to be the merger time inferred by the IMR analysis of an event. $C_{ij}$ is the autocovariance matrix of the noise estimated using the Welch method \cite{welch1967}, as is calculated in other ringdown pipelines \cite{carullo2019, isi2021a, siegel2025}. Note, especially, that the filter does not depend on the QNM amplitude(s) and phase(s), so it does not sample over them but also cannot infer their values. 
See Appendix~\ref{sec:semi-Bayesian} for detailed discussion of the difference between the QNM rational filter and other time-domain analyses.

We define a hypothesis $\mathcal{H}$ as the set of ringdown modes present in a signal at an analysis starting time. The evidence for a hypothesis is given by marginalizing the likelihood over the parameters $M_f$ and $\chi_f$:
\begin{align}
    & \mathcal{Z}(d | \mathcal{H})  =  \notag \\
    & \sum_{M_f,\chi_f} \mathcal{L}(d | M_f, \chi_f, \mathcal{F}_\text{tot}, \Delta t_0) \Pi (M_f, \chi_f) \Delta M_f \Delta \chi_f \, , \label{eq:evidence}
\end{align}
where $\Pi (M_f, \chi_f)$ is the prior for $M_f$ and $\chi_f$, and we have evaluated the integral using a sum with a resolution of ($\Delta M_f, \Delta \chi_f$). Throughout this work, we use a uniform prior over mass and spin. Given an observation, we define a detection statistic ($\mathcal{D}$) used for hypothesis testing to quantify how well an observation prefers a hypothesis $\mathcal{H}$ over an alternative $\mathcal{H}'$:
\begin{equation}
\mathcal{D}(\mathcal{H}:\mathcal{H}') = \log_{10}\frac{\mathcal{Z}(d | \mathcal{H})}{\mathcal{Z}(d | \mathcal{H'})} \, .\label{eq:BF}
\end{equation}
Here $\mathcal{D}$ is analogous to a log Bayes factor but formally differs from the Bayes factors computed by other time-domain ringdown analyses. As such, we refer to our method as a hybrid Bayesian-like analysis, where the parameter estimation of $M_f$ and $\chi_f$ uses a Bayesian framework but the detection of modes with the detection statistic is not strictly Bayesian (for details see Appendix \ref{sec:semi-Bayesian}). We label a hypothesis by its QNM mode content, e.g. $\{220, 221\}$. For a hypothesis, the joint posterior quantile, i.e. the credible contour of which a given set of mass and spin ($M_0, \chi_0$) falls on, can be computed by integrating the likelihoods returned by the rational filter:
\begin{align}
    p(M_{0},\chi_{0})=\frac{\sum_{\mathcal{L}(d|M_f,\chi_f)>\mathcal{L}(d | M_0,\chi_0)}\mathcal{L}(d|M_f,\chi_f)}{\sum_{M_f, \chi_f}\mathcal{L}(d|M_f,\chi_f)} \, . \label{eq:posterior_quantile}
\end{align}
Where we have omitted the $\mathcal{F}_{\rm{tot}}$ and $\Delta t_0$ parameters in the likelihood for compactness and a lower $p(M_{0},\chi_{0})$ value indicates a better match between $(M_{0},\chi_{0})$ and the estimated $(M_f,\chi_f)$.

Finally, the matched-filter SNR of an observation is:
\begin{equation}
\text{SNR} = \frac{\langle h_t | d_t \rangle}{\sqrt{\langle h_t | h_t \rangle}} \, , \label{eq:SNR_MF}
\end{equation}
with
\begin{gather}
\langle h_t | d_t \rangle = \sum_{i,j > t_0} h_i C_{ij}^{-1} d_j \, \label{eq:inner_prod}.
\end{gather}
Note that the matched-filter SNR differs from the optimal SNR, which assumes no additive noise in the data and is given by $\sqrt{\langle h_t | h_t \rangle}$. Throughout the paper, we use SNR to denote the matched filter SNR of a signal for conciseness. 

\section{False-alarm probability and $\mathcal{D}$ threshold}
\label{sec:FAP}
For real GW events, detector noise can mimic ringdown signals and lead to a positive $\mathcal{D}$ when fitting with additional ringdown modes. In the low SNR regime, where the $\mathcal{D}$ for subdominant modes are not large, it can be difficult to determine whether a positive $\mathcal{D}$ arises due to the presence of signal or because of noise fluctuations. We, therefore, quantify the probability that a given $\mathcal{D}$ occurs due to noise using a frequentist approach, empirically determining the distribution of $\mathcal{D}$ that arise from noise. In Sec.~\ref{sec:1_mode}, we use a toy model to describe the procedure for detecting a single ringdown mode. In Sec.~\ref{sec:2_mode}, we discuss a more practical scenario about the detection of a secondary ringdown mode after the dominant mode has already been identified. In Sec.~\ref{sec:improving_detectability}, we discuss modified procedures that can improve the detectability of modes by using knowledge of the remnant BH mass and spin obtained from the IMR analysis. 

\subsection{Single mode detection}
\label{sec:1_mode}
We first consider a simplified scenario and show how we would claim the detection of a single ringdown mode. While it is not a real challenge to detect a dominant ringdown mode in GW observations, we use this relatively simple case to illustrate the procedure of computing the FAP for a single-mode detection. We use data containing only pure noise and compute the $\mathcal{D}$ between $\mathcal{H}$, a \{220\} mode hypothesis, and $\mathcal{H}'$, the null hypothesis (no filter applied). I.e., we compute $\mathcal{D}$ between an overfiltered hypothesis and the correct hypothesis. By doing so, we empirically find the distribution of $\mathcal{D}$, which, if positive, falsely prefers the existence of a 220 mode. We compute $\mathcal{D}$ over the prior space described in Table~\ref{table:prior_params}. The likelihood function depends only on the remnant mass and spin, and we compute it over the entire parameter space without relying on Markov Chain Monte Carlo (MCMC) or nested sampling algorithms. When computing the likelihood over the prior space, $\Delta M_f$ and $\Delta \chi_f$ are the grid spacing between points in $M_f$ and $\chi_f$, respectively. 

We perform the analysis using both a single detector, and two detectors with independent noise. We generate simulated colored Gaussian noise at the Advanced LIGO (aLIGO) design sensitivity~\cite{collaboration2015}.\footnote{The amplitude spectral density is given by \texttt{aligo\_O4high.txt} available at \cite{LVK_noise_curves}.} We generate a 16-s pure-noise data stretch sampled at 4096~Hz and estimate the power spectral density (PSD) using the Welch method \cite{welch1967}. In this pure-noise study, we analyze a 0.2-s segment of data from a start time halfway through the 16-s data stretch (i.e., 8 s after the beginning of the data stretch), searching for a 220 ringdown mode. We perform the analysis for 200 realizations of noise, after which the analysis yields a stabilized distribution of $\mathcal{D}$ (see Appendix \ref{sec:convergence_BF_thresh}). Throughout the rest of the paper, we refer to the procedure described here as the ``$\mathcal{H}$:$\mathcal{H}'$ background study.''
% We check this segment length and sampling rate are sufficient for the analysis, see Appendix \ref{sec:segment_length_and_srate} for the further details. 

\begin{figure}[htb]
\renewcommand\arraystretch{1.4}
\begin{tabular}{ | c | c | c | c |}
\hline
$M_f$ prior & $\Delta M_f$ & $\chi_f$ prior & $\Delta \chi_f$ \\
\hline
[10, 150]\,$M_\odot$ & 0.1$M_\odot$ & [0, 0.99] & 0.005 \\
\hline
\end{tabular}
 \captionof{table}{Prior ranges of $M_f$ and $\chi_f$, and corresponding resolutions, $\Delta M_f$ and $\Delta \chi_f$, used in the analysis.}
\label{table:prior_params}
\end{figure}

\begin{figure}[htb]
\includegraphics[width=\columnwidth]{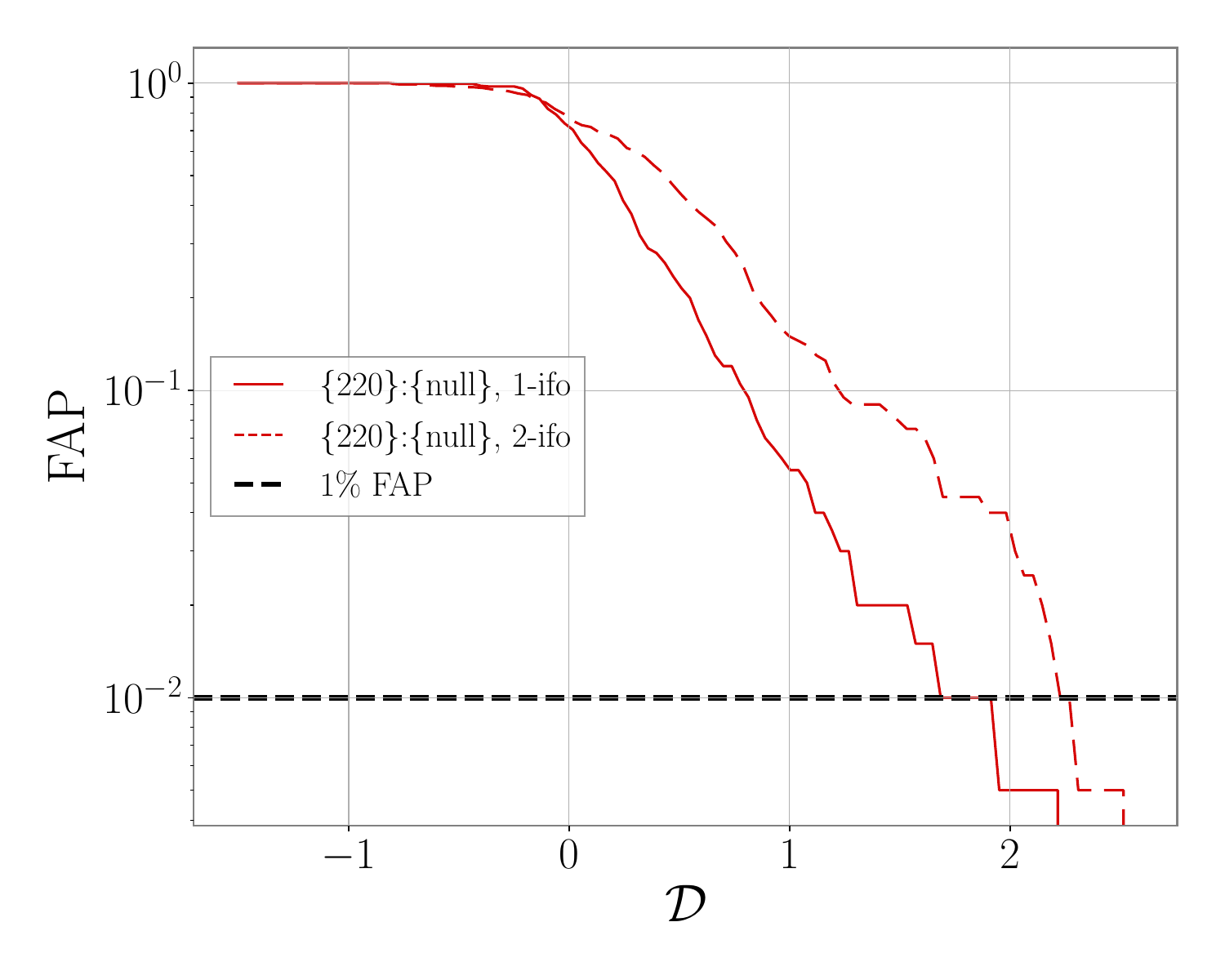}
  \caption{False-alarm probability (FAP) as a function of detection statistic $\mathcal{D}$ when comparing the \{220\} and null hypotheses in pure-noise simulations using a single detector (solid) and two detectors (dotted). Results are obtained from 200 injections in pure colored Gaussian noise at the aLIGO design sensitivity \cite{LVK_noise_curves}.}
 \label{fig:0-mode_injection}
\end{figure}

Figure \ref{fig:0-mode_injection} shows the relationship between FAP and $\mathcal{D}$ when comparing a \{220\} mode hypothesis with the null hypothesis. It shows that many of the simulations recover a positive $\mathcal{D}$, which indicates false positives for the presence of a ringdown mode. We define a threshold on $\mathcal{D}$ for a statistical significance corresponding to a 1\% FAP, denoted by $\mathcal{D}^{\rm thr}$. The thresholds for a single detector (two detectors) is $\mathcal{D}_{\rm \{220\}:\{null\}}^{\rm thr, 1-ifo}=1.68$ ($\mathcal{D}_{\rm \{220\}:\{null\}}^{\rm thr, \, 2-ifo}=2.20$). The two-detector case has a higher threshold because analyzing additional noise from an extra detector increases the probability that statistical fluctuations mimic a QNM signal. The filter is only partially coherent between detectors; signals in different detectors are filtered with coherent real frequencies and damping times, but the filtering does not enforce coherent amplitudes and phases of the ringdown modes. Thus, noise with incoherent amplitudes and phases in two detectors is not intrinsically penalized in the analysis, leading to an increased threshold. For the rest of the paper, we compute the two-detector $\mathcal{D}^{\rm thr, \, 2-ifo}$ by default unless noted otherwise. 
%This makes it likelier that the statistically uncorrelated noise between detectors still increases the BF and explains the higher $\Lambda$. 
For real events, since GW detector noise is non-stationary over long timescales, we recompute the threshold for each observed GW event using the detector noise close to the event ($\mathcal{O}$(hour) before and after the event). When computing the threshold, we ensure the analysis settings (sampling rate, segment length, etc.) are the same as those used when analyzing the event itself. 

To verify that astrophysical ringdown signals are detectable with a 1\% FAP threshold, we estimate the minimum required SNR for a ringdown signal to yield $\mathcal{D} > \mathcal{D}_{\rm \{220\}:\{null\}}^{\rm thr}$. 
%We use the methods introduced later in Sec.~\ref{sec:SNR} and Eq.~\eqref{eq:SNR_vs_BF} and neglect the Occam's term (which tends to be small and always leads to a positive contribution). 
We find that the SNR of a single mode required for an observation above threshold is SNR$_{\rm 1-ifo} = 2.8$ and SNR$_{\rm 2-ifo} = 3.2$ (see Sec.~\ref{sec:SNR} for details). Many existing GW observations have ringdown SNRs above such levels (e.g. $\sim 14$ for GW150914 \cite{isi2019}). Thus, the dominant mode can be confidently detected in GW events.
%and suggests that the BF thresholds are not overly stringent as astrophysical ringdown signals from detected GW events should still be identifiable.  

\begin{figure}[htb]
\includegraphics[width=\columnwidth,clip=true]{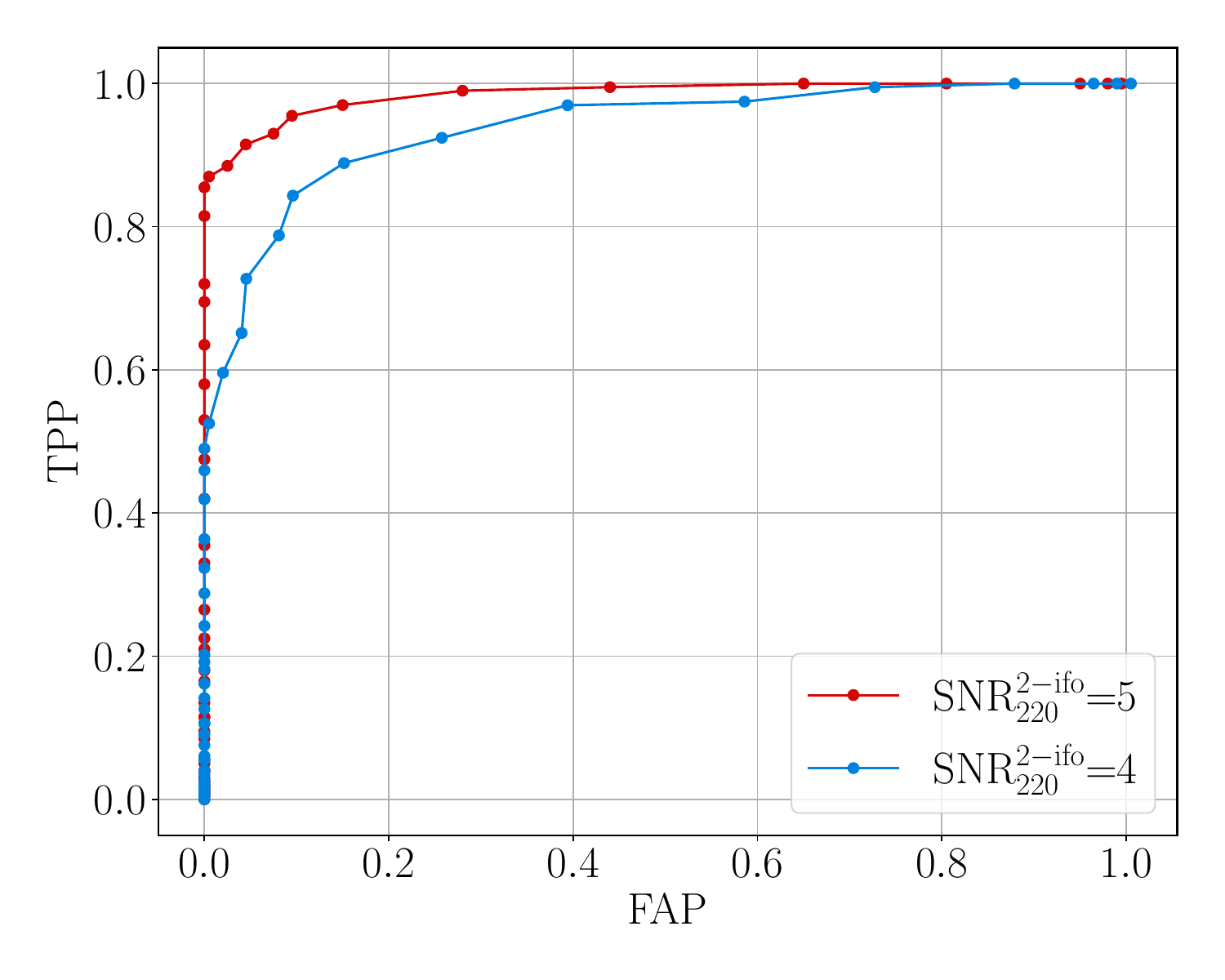}
  \caption{ROC curves for \{220\} mode identification at two different SNR values. 
  %Note that these curves are only illustrative as they depend on the injected remnant spin and QNM phase, even if the injected mode and SNR are fixed.
  }
 \label{fig:1mode_ROC_curve}
\end{figure}

We further evaluate the performance of the method using a receiver operating characteristic (ROC) curve shown in Fig.~\ref{fig:1mode_ROC_curve}, plotting the true-positive probability (TPP) against the FAP. To compute the TPP, we inject a signal with the 220 mode and ringdown parameters consistent with GW150914, given in Table~\ref{table:220_params}. We set the mode amplitude to two different values, resulting in SNR$^{\rm 2-ifo} = 4$ and $5$, to evaluate the recovery rate. For each choice of the mode amplitude, the waveform is injected into 200 different noise realizations. Then, we compute $\mathcal{D}_{\rm\{220\}:\{null\}}$ for each injection and plot the TPP for each given FAP. 
%In this case we expect that positive $\mathcal{D}$ are true-positive caused by the signals and are not produced by noise fluctuations. 
%We compare the TPP for a given $\mathcal{D}$ with the two-detector FAP results in Fig.~\ref{fig:0-mode_injection} where instead positive $\mathcal{D}$ are false-alarms. 
%If we start with a high BF threshold, the FAP is low, but the true-positive probability (TPP) is also low. 
%As the BF threshold decreases, both TPP and FAP increase. An ideal BF threshold approaches a Heaviside step function with 100\% true-positive probability with a 0\% FAP so can always accurately differentiate between true-positives and false-alarms. 
Fig.~\ref{fig:1mode_ROC_curve} shows that, as expected, signals with a larger SNR are easier to recover. At both of these relatively low SNRs, the signal can achieve TPP $\gtrsim 90\%$ for FAP $\gtrsim 20\%$, demonstrating that the rational filter can efficiently distinguish the signal from noise. Note, however, that the ROC curve depends on the injection parameters (remnant BH mass, spin, and QNM phase) even when the SNR is fixed. Therefore, these results should only be interpreted as indicative. 

\begin{figure}[htb]
\renewcommand\arraystretch{1.4}
\begin{tabular}{ | c | c | c |}
\hline
$\phi_{220}$ & $M_{\rm inj}$ &  $\chi_{\rm inj}$ \\
\hline
4.81 & 68.5$M_\odot$ & 0.692 \\
\hline
\end{tabular}
  \captionof{table}{Parameters of the \{220\} waveform injected to calculate the ROC curve shown in Fig.~\ref{fig:1mode_ROC_curve}. These ringdown parameters are consistent with GW150914 \cite{lovelace2016, clarke2024}.}
 \label{table:220_params}
\end{figure}

\subsection{Identifying a second mode}
\label{sec:2_mode}
We now extend this method to detecting a subdominant mode, assuming the dominant mode has been confidently identified. Conceptually, the idea is now to inject a single QNM into noise and compute $\mathcal{D}$ between a hypothesis with two QNMs and the hypothesis with the single correct QNM. 
We choose to inject a set of controlled QNMs instead of full GW waveforms to avoid contamination from additional mode contents, e.g., overtones, that can impact the background estimate.
We again follow the procedure described in Sec.~\ref{sec:1_mode} to create 16~s of Gaussian noise, colored with the aLIGO design PSD and sampled at 4096~Hz. We inject a ``ringup-ringdown'' QNM signal into the generated noise:
\begin{gather}
    x(t) = h(|t-8~{\rm s}|) \, , \label{eq:ringup_ringdown}
\end{gather}
where $|\cdot|$ denotes the absolute value function, $h(t)$ is given in Eq.~\eqref{eq:QNM_def}, and we include the ringup component to avoid the impact of spectral leakage. The QNM is injected with a start time halfway through the timeseries (i.e., 8~s after the start). Here we always inject the 220 mode, which is expected to be dominant for quasi-circular binary BH coalescences, but the technique is readily applicable if a different QNM is dominant. The parameters of the injected QNM are randomly sampled from the ranges given in Table~\ref{table:signal_params}. Note that the complex frequencies of the injected QNMs are determined by the properties of the remnant BHs, where $M_{\rm inj}$ and $\chi_{\rm inj}$ are the mass and spin of the remnant, respectively. We inject a range of SNRs, but the computed background threshold does not depend on the SNR of the injection because the injected mode is always completely filtered out at the maximum likelihood values (which is the dominant term of the computed evidence) for both hypotheses we consider. We start our analysis 3~$M_{\rm inj}$ after the QNM start time to ensure no impact is introduced to the analysis by the ringup part of the injection, but the computed background threshold is insensitive to the start time of the analysis (for $\Delta t_0 \geq 0$). We analyze 0.2~s of data as a standard setup. For real GW events, the injected signal parameters for the background study could be adjusted to better represent the ranges that are relevant to the true event, and the segment length and downsampling effects would be verified to not affect the estimated background threshold \cite{siegel2023}.
%if preliminary analysis with the rational filter suggests the remnant properties do not lie within the ranges in Table~\ref{table:signal_params}. 

\begin{figure}[htb]
\renewcommand\arraystretch{1.4}
\begin{tabular}{ | c | c | c | c |}
\hline
$M_{\rm inj}$ & $\chi_{\rm inj}$ & Mode phase & SNR \\
\hline
[30, 120] $M_\odot$ & [0, 0.95] & [0, $2\pi$] & [4, 200] \\
\hline
\end{tabular}
  \captionof{table}{Ranges of injected signal parameters when computing the threshold in Fig.~\ref{fig:1-mode_injection}}
 \label{table:signal_params}
\end{figure}

\begin{figure}[htb]
\includegraphics[width=\columnwidth,clip=true]{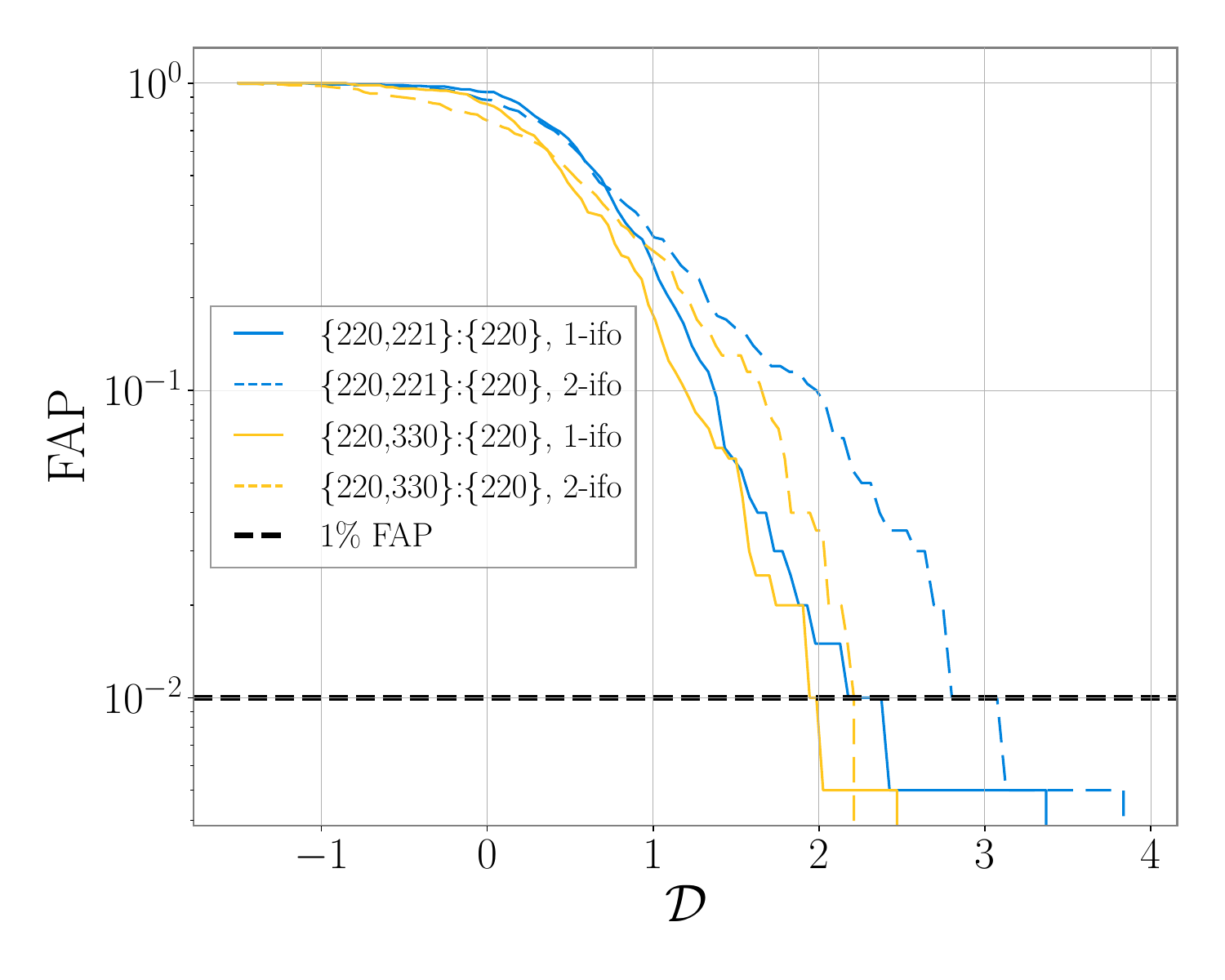}
\caption{FAP as a function of $\mathcal{D}$ when comparing the \{220,221\}:\{220\} (blue) and \{220,330\}:\{220\} (yellow) for injections of \{220\}-only signals.  Solid and dotted curves are for a single detector and two detectors, respectively. The results are based on 200 simulations in colored Gaussian noise at aLIGO design sensitivity.}
 \label{fig:1-mode_injection}
\end{figure}

We compute $\mathcal{D}$ for a particular two-mode hypothesis (e.g., \{220,221\} or \{220,330\}) compared to the correct \{220\} hypothesis. We use the same prior region as those in Table~\ref{table:prior_params} and analyze 200 injections to determine the empirical $\mathcal{D}$ threshold. Fig.~\ref{fig:1-mode_injection} shows the results. They  are qualitatively similar to the pure-noise case (Sec.~\ref{sec:1_mode}) where the majority of injections recover positive $\mathcal{D}$ values, and the two-detector threshold ($\mathcal{D}_{\rm \{220,221\}:\{220\}}^{\rm thr, \, 2-ifo}=2.79$) is larger than the single-detector one ($\mathcal{D}_{\rm \{220,221\}:\{220\}}^{\rm thr, 1-ifo}=2.17$). The $\mathcal{D}$ thresholds also depend on the identity of the secondary mode. In fact, when overfiltering the same set of \{220\}-only injections with the \{220,330\} and \{220,221\} models, the thresholds are $\mathcal{D}_{\rm \{220,330\}:\{220\}}^{\rm thr, \, 2-ifo}=2.17$ and $\mathcal{D}_{\rm \{220,221\}:\{220\}}^{\rm thr, \, 2-ifo}=2.79$, respectively. This is because the distinguishability of different modes depend on how close their complex frequencies are. The $\mathcal{D}^{\rm thr}$, therefore, needs to be computed for each specific secondary mode being tested. 

\begin{figure}[htb]
\includegraphics[width=\columnwidth,clip=true]{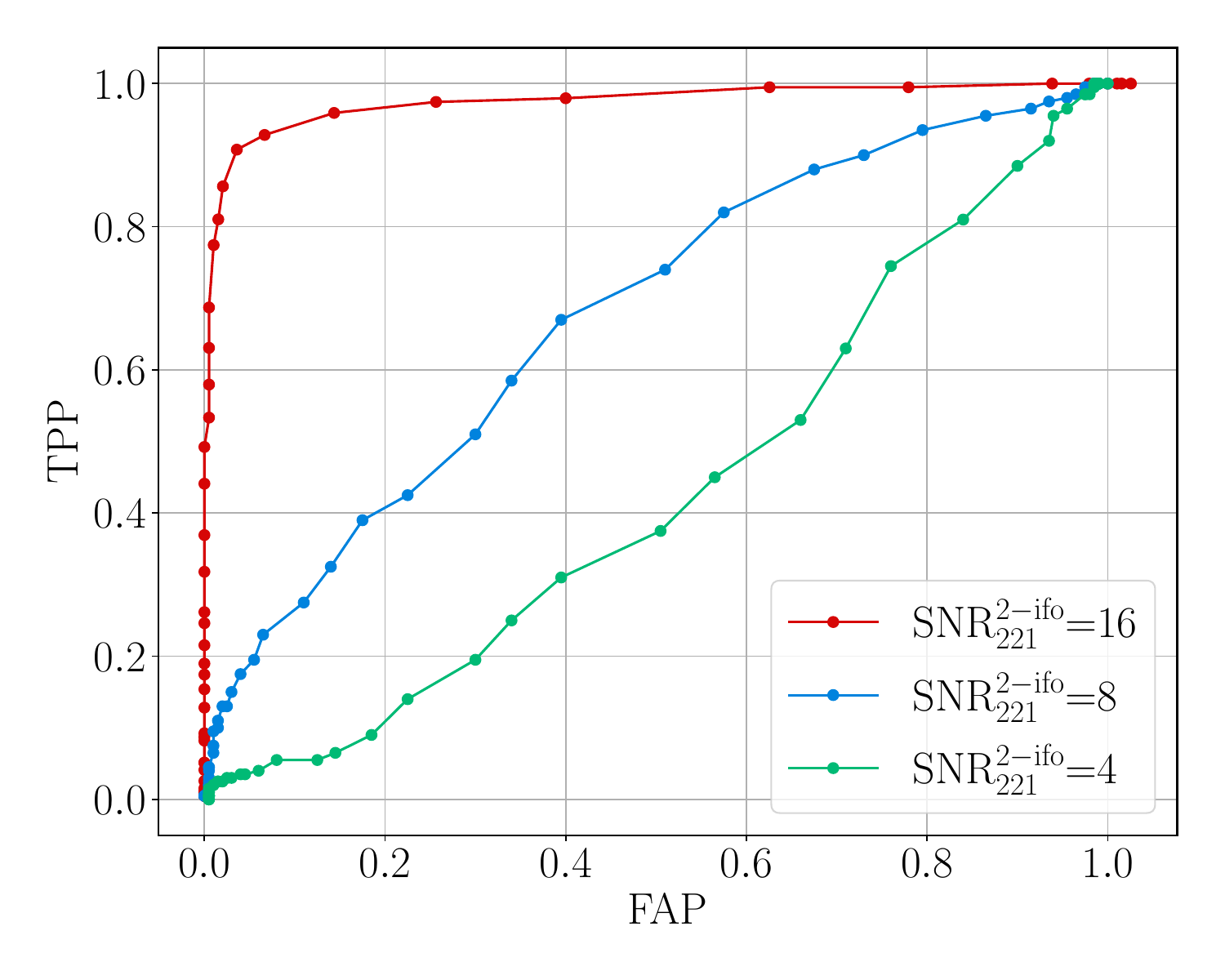}
  \caption{ROC curves when comparing the the true \{220,221\} hypothesis to the single-mode \{220\} hypothesis with three different SNRs of the 221 mode. 
  %Note that these are only illustrative and the exact behavior will depend on the injection parameters even if using the same SNRs or mode combinations.
  }
 \label{fig:2mode_ROC_curve}
\end{figure}

Following the approach in Sec.~\ref{sec:1_mode}, we construct the ROC curve for the two-mode scenario to quantify the sensitivity and robustness of our method. We inject a \{220,221\} signal with parameters given in Table~\ref{table:220+221_params}, consistent with GW150914, with the waveform multiplied by different scale factors so that the SNR of the subdominant mode is SNR$_{221}$ = 4, 8, 16. Each waveform is injected into 200 independent noise realizations to compute the TPP. As shown in Fig.~\ref{fig:2mode_ROC_curve}, the ROC curves for SNR$_{221}=4$ and $8$ do not show ideal recovery. However, signals with SNR$_{221}=16$ are confidently detectable. 

%Note, this ROC curve is only illustrative as the sensitivity of the rational filter depends on the injection parameters of the signal (remnant mass, spin, and QNM phase) even when controlling for the SNR and injected set of modes. 

\begin{figure}[htb]
\renewcommand\arraystretch{1.4}
\begin{tabular}{ | c | c | c | c | c | c | c | c |}
\hline
$A_{220}$ & $A_{221}$ & $\phi_{220}$ & $\phi_{221}$ & $M_{\rm inj}$ &  $\chi_{\rm inj}$ \\
\hline
1.0 & 4.24 & 4.81 & 0.676 & 68.5 & 0.692 \\
\hline
\end{tabular}
  \captionof{table}{Synthetic signal parameters of the two-mode \{220,221\} ringdown waveform (GW150914-like) used in simulations to calculate the ROC curve in Fig.~\ref{fig:2mode_ROC_curve}.}
 \label{table:220+221_params}
\end{figure}

\subsection{IMR-conditional injections}% Improving mode detectability}
\label{sec:improving_detectability}
The background study described in Secs.~\ref{sec:1_mode} and \ref{sec:2_mode} does not rely on the IMR analysis of the event. Any ringdown mode surpassing the threshold as described above can be tested for consistency with the remnant BH parameters obtained independently from the IMR parameter estimation (PE). In this section, however, we introduce techniques that can improve the sensitivity of the rational filter to subdominant modes by including the information from the IMR PE. These techniques could be useful for current GW observations where the ringdown SNRs are generally not high. 

%The robustness of the method in Sec.~\ref{sec:1_mode} and~\ref{sec:2_mode}  arises from two factors: the use of agnostic injection parameters and the use of a full prior. The agnostic injection parameters imply that injections are randomly sampled over a range informed by the analysis of the GW event with the rational filter. 
%These injections do not use any of the information from the IMR analysis. 
%An alternative way of injecting signals conditional on the correctness of the IMR parameter estimation is described in Sec.~\ref{sec:conditional_full}. It is also possible to restrict the prior range (i.e. that given in Table~\ref{table:prior_params}) to remove features of the likelihood such as bimodality, this method is discussed in Sec.~\ref{sec:conditional_restricted}. 

\subsubsection{IMR-conditional injections with full prior}
\label{sec:conditional_full}
When evaluating the background, we have two options of selecting injection parameters: 1) randomize the $M_{\rm inj}$ and $\chi_{\rm inj}$ in the full prior range, as described above, and 2) sample $M_{\rm inj}$ and $\chi_{\rm inj}$ from the posterior obtained from the IMR analysis. In both cases the amplitude and phase of the QNMs injected are randomized. These two options correspond to two different motivations: a) \emph{IMR-agnostic} analyses for IMR consistency tests or independent tests of GR via BH spectroscopy, and b) \emph{IMR-conditional} studies, assuming that the IMR PE results are reliable and aiming for enhanced sensitivity to identify subdominant modes, which are useful for investigating initial conditions of the binary \cite{zhu2023}. 

\begin{figure}[htb]
\includegraphics[width=\columnwidth,clip=true]{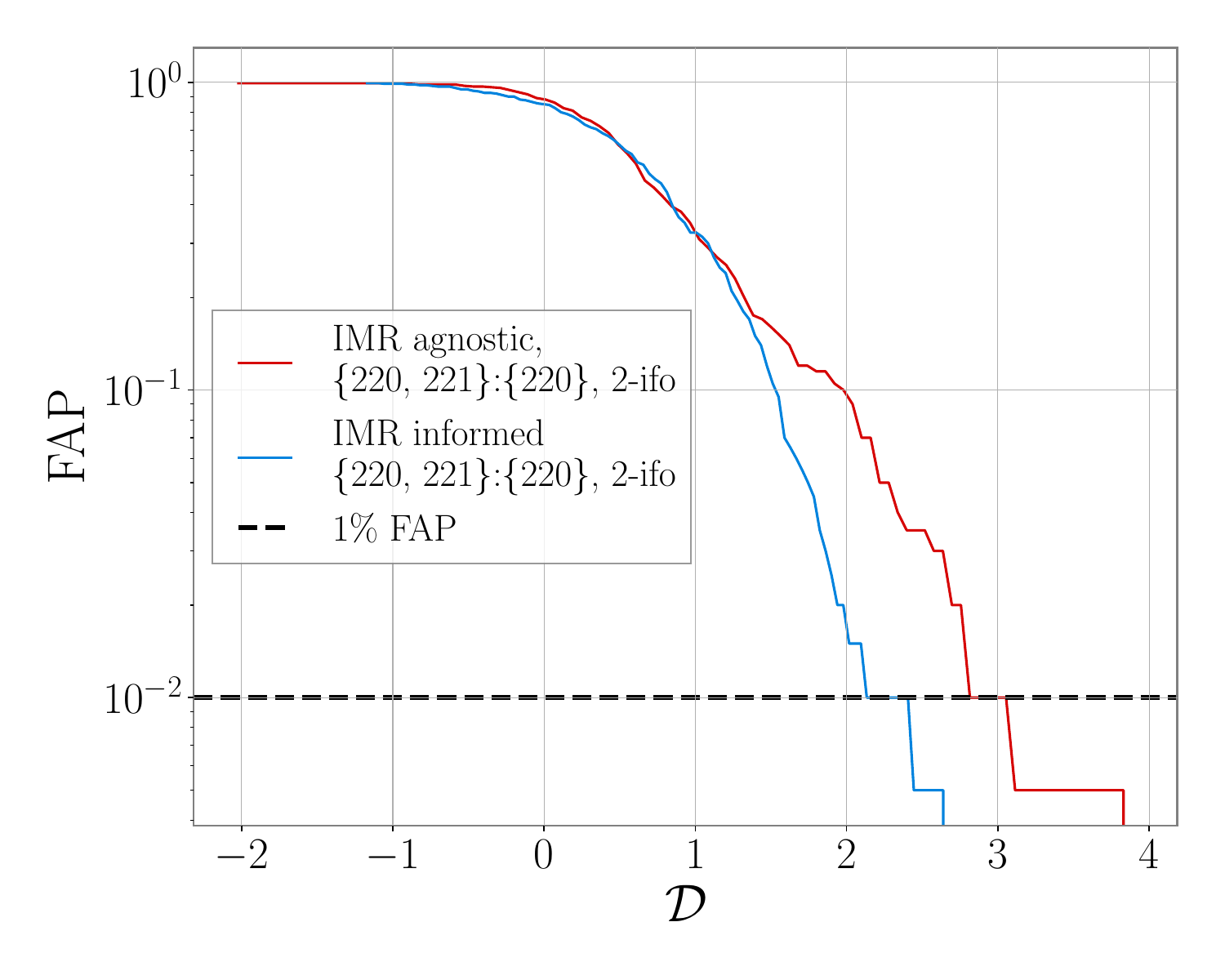}
  \caption{FAP as a function of $\mathcal{D}$ when comparing the \{220,221\}:\{220\} hypotheses using IMR-agnostic and IMR-conditional injections of \{220\}-only signals. The parameters for the conditional injections are sampled from the posterior distribution of GW150914 inferred from IMR \cite{theligoscientificcollaborationandthevirgocollaboration2024}. For each injection configuration, the results are obtained from 200 synthetic signals injected into colored Gaussian noise at aLIGO design sensitivity in two detectors.}
 \label{fig:agnostic_vs_conditional_injection}
\end{figure}

To study how conditional injections impact the background distribution and the threshold, we repeat the analysis in Sec.~\ref{sec:2_mode}---injecting a 220 QNM into a two-detector network and overfiltering with a \{220,221\} mode model. This time, however, the remnant BH mass and spin for the injections are sampled from the posterior distribution of GW150914 inferred from the IMR \cite{theligoscientificcollaborationandthevirgocollaboration2024, collaboration2022}. The analysis still uses the full prior space given in Table~\ref{table:signal_params}. The background distribution from the conditional injections is shown in Fig.~\ref{fig:agnostic_vs_conditional_injection}, where the thresholds for the IMR-agnostic injections and conditional injections are $\mathcal{D}_{\rm \{220,221\}:\{220\}}^{\rm thr, \, agnostic}=2.79$ and $\mathcal{D}_{\rm \{220,221\}:\{220\}}^{\rm thr, \, conditional}=2.11$, respectively. Therefore, using conditional injections for a background study lowers the threshold required for detecting subdominant modes and increases the sensitivity to these modes. 

\subsubsection{IMR-conditional injections with restricted priors}
\label{sec:conditional_restricted}
For certain mode configurations, the likelihoods computed over the whole parameter space contain bimodal features. Specifically, for signals with a 220 fundamental mode and a subdominant 330 or 440 mode, the likelihood contours in the mass and spin plane show bimodality. This is because, for a given $\chi_f$, the damping times of the 220 and 330 (440) are similar, but their real frequencies differ by a ratio of approximately 2:3 (1:2). Since the real frequency of a QNM is scaled by the remnant BH mass, a 220 mode can almost entirely be filtered out by a 330 filter constructed with a similar $\chi_f$ value and a $M_f$ value $\sim 1.5$ times higher than the true $M_f$. 
% \sma{This is not precise. When you change $M_f$, the damping time is also changed. So $\chi_f$ needs to be adjusted accordingly. In fig.~\ref{fig:220+330}, the two peaks don't have similar $\chi_f$.}. 
An example of this bimodality is shown in Fig.~\ref{fig:220+330_full}, where we inject a \{220,330\} waveform with parameters given in Table~\ref{table:220+330_params} (consistent with those predicted for a high-mass-ratio system which excites a non-negligible 330 mode). The signal is injected into colored Gaussian noise, and its amplitude is scaled so that the total ringdown SNR and the 330-mode SNR are SNR$_\text{tot}^{\rm 2-ifo} = 20.5$ and SNR$_{330}^{\rm 2-ifo}=4$, respectively. Fig.~\ref{fig:220+330_full} shows the likelihood contours obtained from the rational filter using the correct \{220,330\} QNM model. The secondary peak is found at a mass approximately 1.5 times higher than the mass corresponding to the main peak. 

\begin{figure}[htb]
\renewcommand\arraystretch{1.4}
\begin{tabular}{ | c | c | c | c | c | c | c | c |}
\hline
$A_{220}$ & $A_{330}$ & $\phi_{220}$ & $\phi_{330}$ & $M_{\rm inj}$ &  $\chi_{\rm inj}$ & SNR$_\text{tot}$ & SNR$_\text{330}$ \\
\hline
1.0 & 0.081 & 0.529 & 1.457 & 72 & 0.64 & 20.5 & 4 \\
\hline
\end{tabular}
  \captionof{table}{Synthetic signal parameters of the two-mode \{220,330\} ringdown waveform. These parameters are consistent with a binary system with where $m_1/m_2 = 10$, $\chi_1 = \chi_2 = 0.5$, where $m_1, m_2$ and $\chi_1, \chi_2$ are the initial masses and spins of the two progenitor BHs, and the system has an inclination angle, $\iota=\pi/2$, and polarization angle $\psi=0$, as given by the \texttt{postmerger} package \cite{pacilio2024}.}
 \label{table:220+330_params}
\end{figure}

If the prior is restricted to $M_f = [30, 85]M_\odot$, $\chi_f = [0.35, 0.95]$, only a single peak appears in the likelihood contours, as shown in Fig.~\ref{fig:220+330_restricted}. The restricted prior is chosen to only include the peak that agrees with the remnant BH parameters used for injections (or IMR posterior if analyzing real GW events). This restricted prior also corresponds to the expectation that the 220 QNM amplitude should be larger than the 330 amplitude, as is the case for quasi-circular binary BH mergers (although see \cite{zhu2023} for precessing binaries). Thus, adopting a restricted prior is not strictly contingent on comparisons with IMR analyses. 
%this restricted prior would also be used when estimating the BF threshold for the 99\% FAP which also demonstrates bimodality. The reduction in the BF of the event itself might therefore be outweighed by the reduction in the BF threshold which would make events with low SNR more detectable. 

\begin{figure}[htb]
\newlength{\imagewidth}
\settowidth{\imagewidth}{\includegraphics{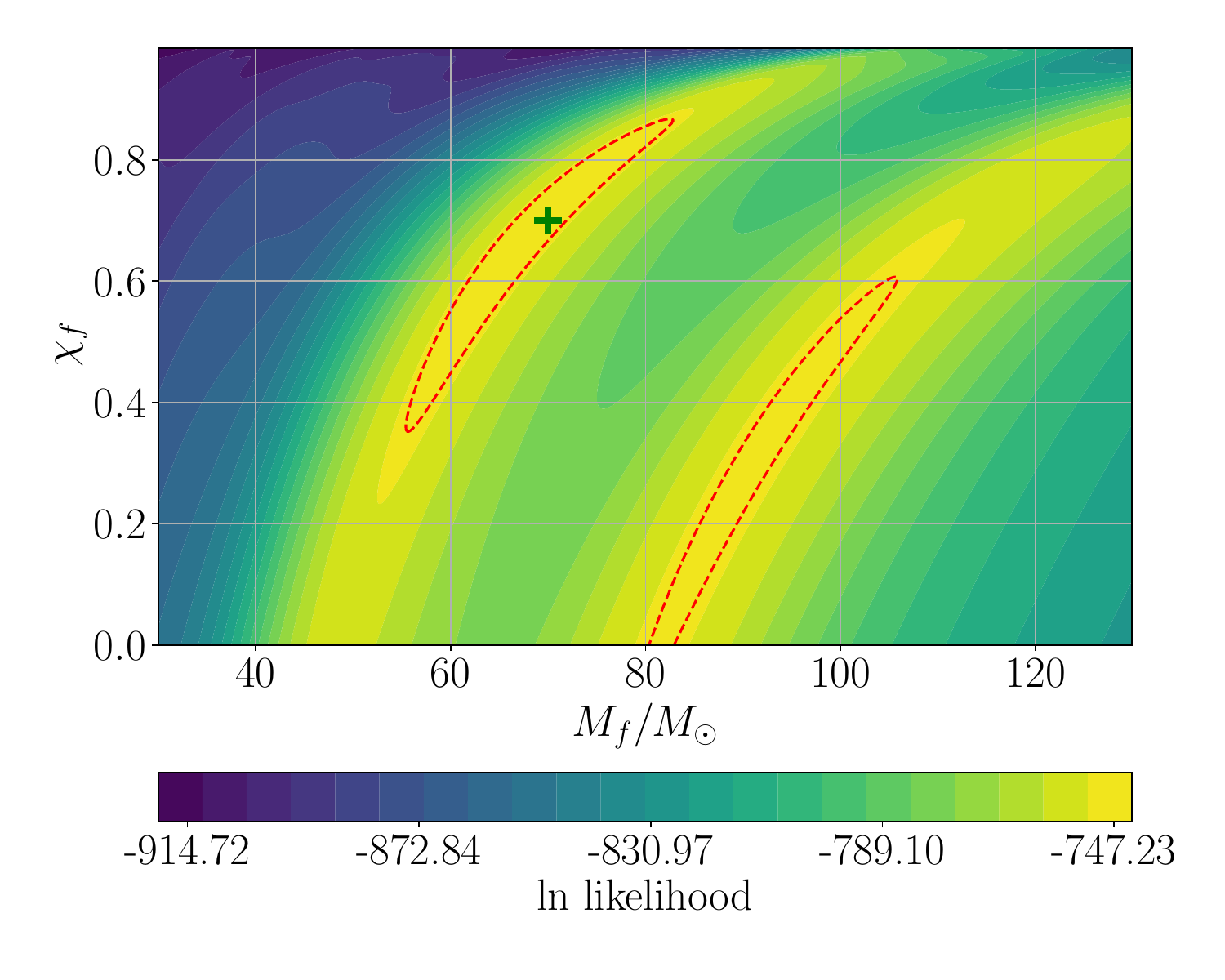}}
    \subfloat[\centering Full prior\label{fig:220+330_full}]{\includegraphics[width=0.5\textwidth]{BF_threshold/220+330_full.pdf}}\hfill
    \subfloat[\centering Restricted prior \label{fig:220+330_restricted}]{\includegraphics[width=0.5\textwidth]{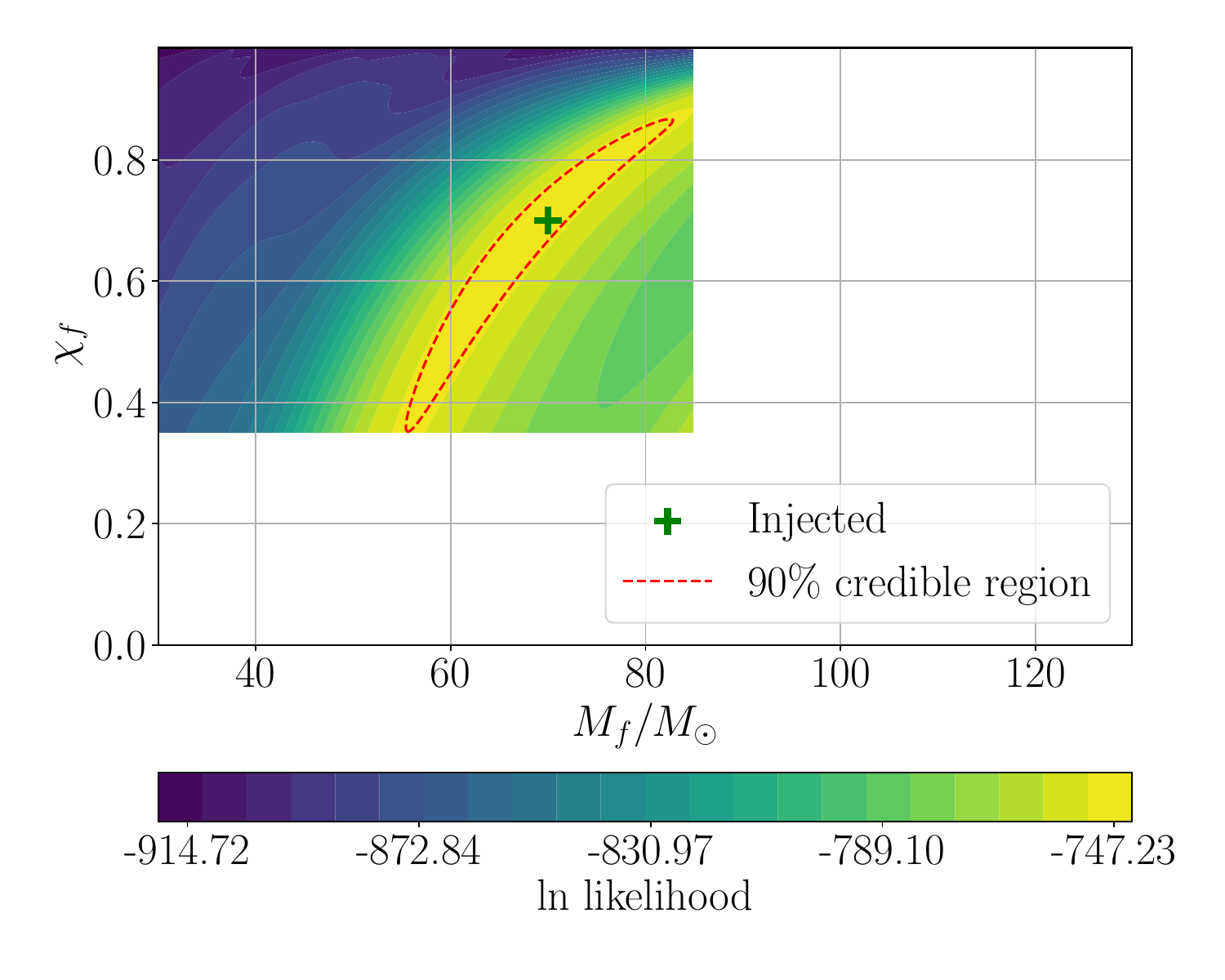}}\hfill
    \caption{Likelihood contours of $M_f$ and $\chi_f$ for a synthetic waveform with \{220,330\} modes obtained using the correct \{220,330\} rational filter. Panels (a) and (b) are for the two choices of the prior range. The green plus mark indicates the true injection parameters given in Table~\ref{table:220+330_params}.}
    \label{fig:220+330}
\end{figure}

\begin{figure}[htb]
\includegraphics[width=\columnwidth,clip=true]{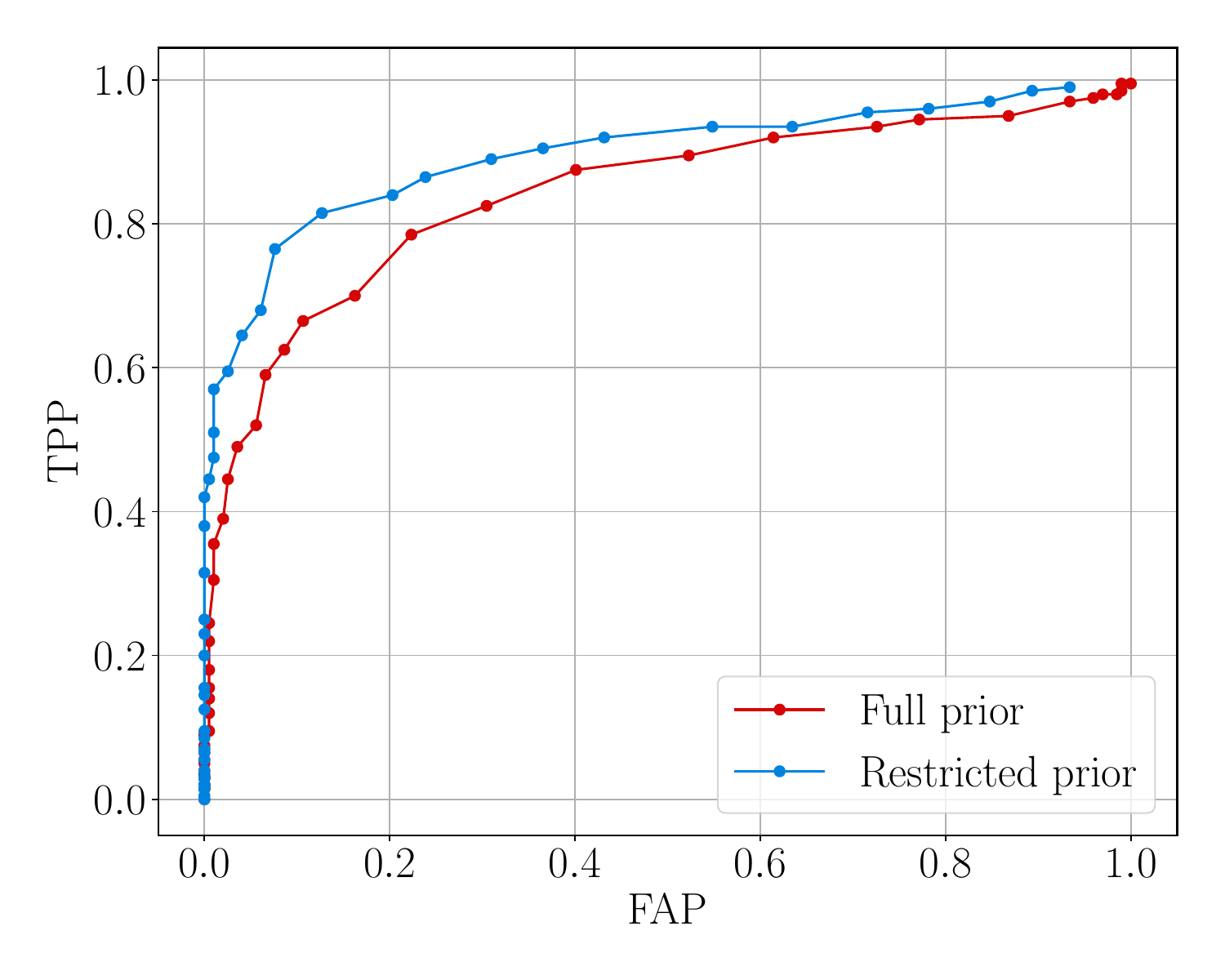}
  \caption{ROC curves for detecting the subdominant 330 mode using the full prior and the restricted prior. The results are obtained from 200 injections in colored Gaussian noise with SNR$_{330}^{\rm 2-ifo}=4$. }
 \label{fig:220+330_ROC_curve}
\end{figure}

Restricting the prior range may reduce the $\mathcal{D}$ of the signal due to the fact that the likelihoods are integrated over the whole prior range. However, the threshold obtained by analyzing the background over the restricted prior range also decreases. To determine whether using the restricted prior improves the detectability of a subdominant 330 mode, we construct a ROC curve for the \{220,330\}:\{220\} analysis, similar to Sec.~\ref{sec:2_mode}. In this case, we compute the TPP from 200 injections of the \{220,330\} waveform in colored Gaussian noise (all simulated with the parameters given in Table~\ref{table:220+330_params}). The threshold is determined by performing a \{220,330\}:\{220\} background study where the 220-only injection parameters are fixed (Table~\ref{table:220+330_params}).
The study is repeated for the two prior choices, full and restricted, and the results are shown in Fig.~\ref{fig:220+330_ROC_curve}. 
The restricted prior improves the sensitivity to the subdominant 330 mode.\footnote{Both ROC curves are better than the SNR$_{221}^{\rm 2-ifo}=4$ case in Fig.~\ref{fig:2mode_ROC_curve} because the 330 QNM is more separated in frequency and damping time from the 220 mode than the 221 mode.} In this analysis, the restricted prior range is chosen such that only the main peak of the likelihood in the mass and spin plane, which agrees with IMR, is included. There are, however, alternative choices for the restricted prior range, and an optimal prior range may exist for a given system that maximizes the detectability of a subdominant mode. We defer the investigation of an optimized choice of priors to future work. In this example, we only demonstrate the application of this technique for waveforms with the \{220,330\} modes. It is, in principle, applicable to any mode combination which shows multi-modality or features in the likelihoods which are a-priori known to be unphysical based on numerical relativity or the IMR analysis of the event. 

%==========================================================================
\section{SNR estimation}
\label{sec:SNR}
In this section, we describe how to use the rational filter to estimate the ringdown SNR after all detectable modes are identified. We also explain the limitations of SNR estimates for individual modes.

A generic relationship for the Bayes factors (BFs) of two hypotheses, $\mathcal{H}$ and $\mathcal{H}$', corresponding to signal templates, $h$ and $h'$, versus the SNR of a signal is \cite{cornish2011}:
\begin{gather}
    \ln (\text{BF }[h':h]) = (1-\text{FF}^2) \frac{\text{SNR}^2}{2} + \Delta \ln \mathcal{O} \, , \label{eq:BF_vs_SNR}
\end{gather}
where FF is the fitting factor, i.e., the best match that can be achieved between the two hypotheses, defined by: 
\begin{gather}
\text{FF} = \text{max} \frac{(h|h')}{\sqrt{(h|h) (h'|h')}} \, . \label{eq:FF}
\end{gather}
The maximum is taken over all the parameters of $h$ and $h'$, and $\mathcal{O}$ is the Occam's factor approximated by:
\begin{gather}
\mathcal{O} = \frac{V_{\text{post}, 95\%}}{V_\text{prior}} \, , \label{eq:occams}
\end{gather}
where $V_\text{post, 95\%}$ is the volume of the 95\% credible region of the posterior, and $V_\text{prior}$ is the volume of the prior. Since $\mathcal{D}$ is analogous to $\ln (\text{BF})$, we should see the same relation in Eq.~\eqref{eq:BF_vs_SNR} between $\mathcal{D}$ and SNR. 

\begin{figure}[htb]
\includegraphics[width=\columnwidth,clip=true]{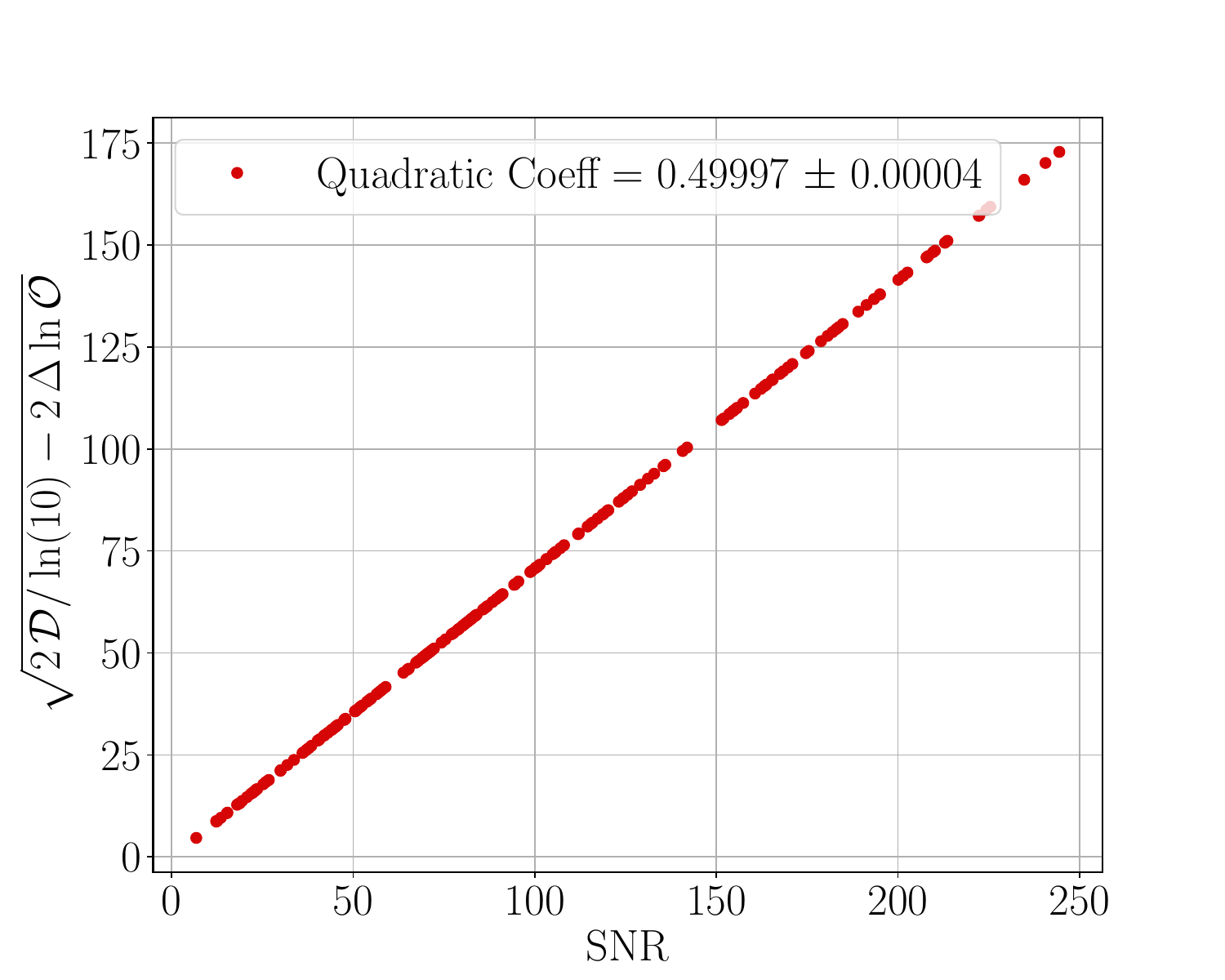}
  \caption{The scaling between $\mathcal{D}$ (comparing the injected mode hypothesis and the null hypothesis) and the SNR of the injected signal. Results are calculated from 200 injections (see Table~\ref{table:signal_params} for injection parameter ranges).  
  %with randomized masses, spins, amplitudes, and phases for the 220 ringdown mode into the aLIGO design ASD.
  }
 \label{fig:BF_scaling_FF=0}
\end{figure}

To verify this we inject signals with \{220\} QNM only into colored Gaussian noise with randomized parameters (remnant BH mass and spin, and mode amplitude and phase) drawn over the same ranges as listed in Table~\ref{table:signal_params}. We compute $\mathcal{D}$ between the correct 220-mode hypothesis and the null hypothesis using the same prior range as that described in Sec.~\ref{sec:FAP}. Since the null hypothesis implies no signals are present (i.e. $h(t) = 0$), we can see from Eq.~\eqref{eq:FF} that FF $=0$. For each injection, we also directly compute the Occam's factor $\mathcal{O}$ from the posterior using Eq.~\eqref{eq:occams}. The results are shown in Fig.~\ref{fig:BF_scaling_FF=0}, we fit the data points with a second-order polynomial and find that the quadratic coefficient is $0.49997 \pm 0.00004$, which agrees precisely with Eq.~\eqref{eq:BF_vs_SNR} when FF $=0$. 

\begin{figure}[htb]
\includegraphics[width=\columnwidth,clip=true]{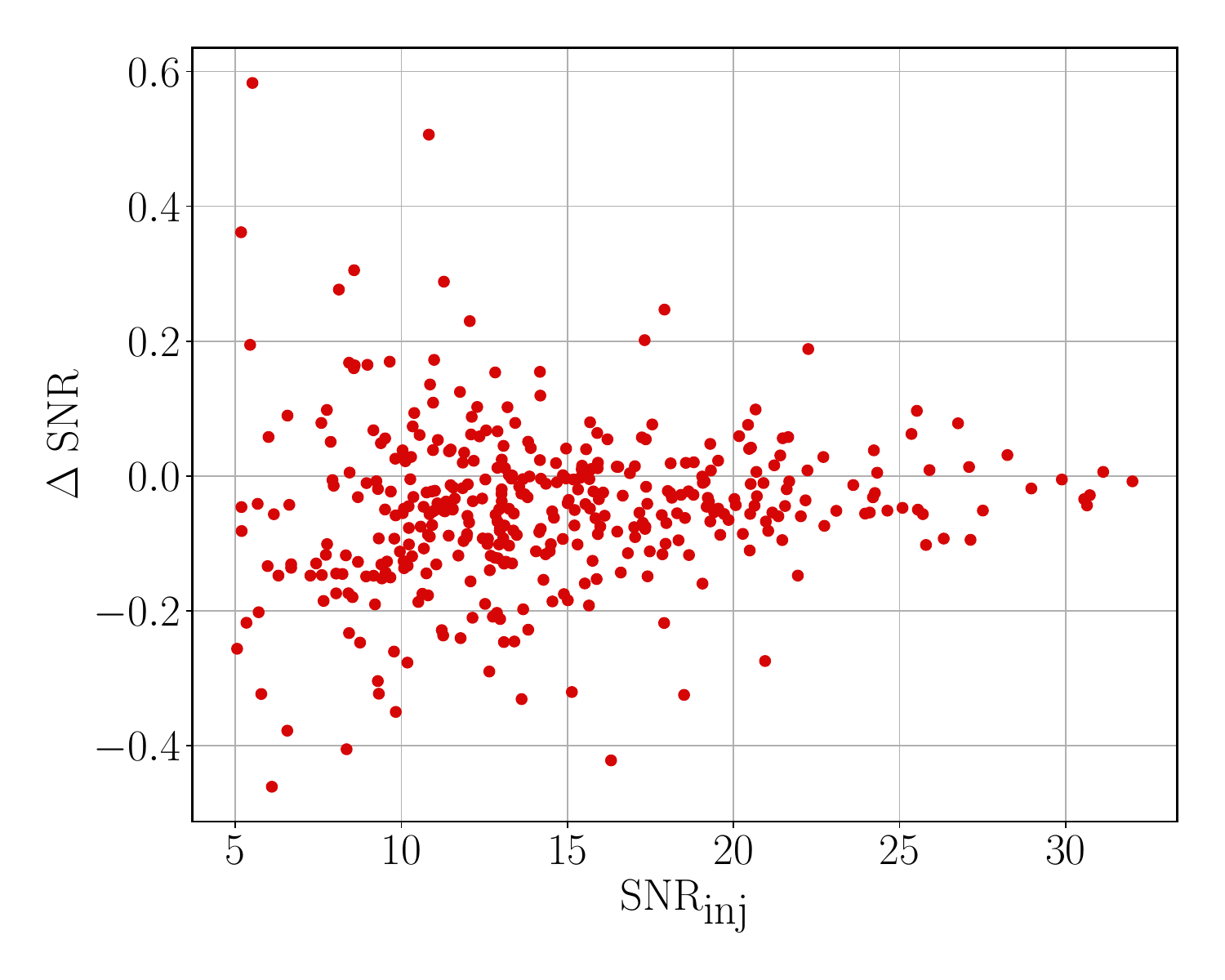}
  \caption{Error in the estimated SNR versus the true SNR for 200 injections of a \{220\}-only ringdown signal in colored Gaussian noise at the aLIGO design sensitivity.}
 \label{fig:SNR_error}
\end{figure}

We now use Eq.~\eqref{eq:BF_vs_SNR} when FF $=0$ to find an expression for SNR:
\begin{gather}
\text{SNR} = \sqrt{2 \,\mathcal{D}/\ln(10) - 2\,\Delta \ln \mathcal{O}}. \label{eq:SNR_vs_BF}
\end{gather}
%For the analysis of a GW event, we can measure the BF and Occams factor of a ringdown template compared to the null hypothesis and so can calculate the matched-filter SNR. 
To quantify the accuracy of this estimation, we consider the error in the estimated SNR for an injection:
\begin{gather}
    \Delta \,\text{SNR} = \text{SNR}_\text{inj} - \text{SNR} \label{eq:Delta_SNR} \, ,
\end{gather}
where $\text{SNR}_\text{inj}$ is the matched filter SNR of the injected signal.
We inject ringdown signals with ${\rm SNR}_\text{inj} \in [5,35]$, which is consistent with the expected SNR range for current-generation GW detectors, and compute the recovered $\Delta \, \text{SNR}$. Fig.~\ref{fig:SNR_error} shows the result and demonstrates that the computed SNRs are broadly reliable with $|\Delta \text{SNR}| \lesssim 0.6$ and the error further decreases for signals with larger SNRs. The error arises because Eq.~\eqref{eq:BF_vs_SNR} is derived using Laplace's approximation for the likelihood distribution \cite{cornish2011, romano2017}. Particularly, the definition of Occam's factor in Eq.~\eqref{eq:occams} assumes that the likelihoods are Gaussian distributed and peaked at the maximum likelihood value. However, for the rational filter, this assumption is generally not true, which leads to residual errors in the estimated SNR. This effect is more significant in the low-SNR regime where the Occam's factor is larger, but we still find that the SNR estimation is sufficiently accurate. 
%and where Fig.~\ref{fig:SNR_error} shows that there is a small systematic bias to overestimate the SNR. The overestimated SNR is due to extra power from the noise captured by the rational filter due to slight deviations between the estimated and real PSD. 
%Despite this, we numerically find $\Delta \text{SNR}<0.6$ for all injections with ${\rm SNRs} >9$. 
We repeat this verification, that $\Delta \text{SNR}\lesssim 1$, for injections of up to three modes in noise generated with different PSDs and find the result remains robust.

This technique can compute the total SNR of an arbitrary number of identified modes, but cannot estimate the SNR of individual modes when the signal contains more than one mode. This is because the FF in Eq.~\eqref{eq:BF_vs_SNR} between non-trivial hypotheses cannot be calculated. 
%so Eq.~\eqref{eq:BF_vs_SNR} has two unknowns (FF and SNR) and cannot be solved. 
See Appendix~\ref{sec:2nd_mode_SNR} for a more detailed discussion. 

%==========================================================================
\section{Analysis procedure}
\label{sec:analysis_procedure}
In this section, we describe the analysis procedure for analyzing GW events and identifying their mode composition. We discuss how to identify the correct QNM if multiple subdominant modes are potential candidates (Sec.~\ref{sec:identification}) and describe the complete workflow (Sec.~\ref{sec:workflow}).

%==========================================================================
\subsection{Identify the correct mode}
\label{sec:identification}
\begin{figure}[htb]
\settowidth{\imagewidth}{\includegraphics{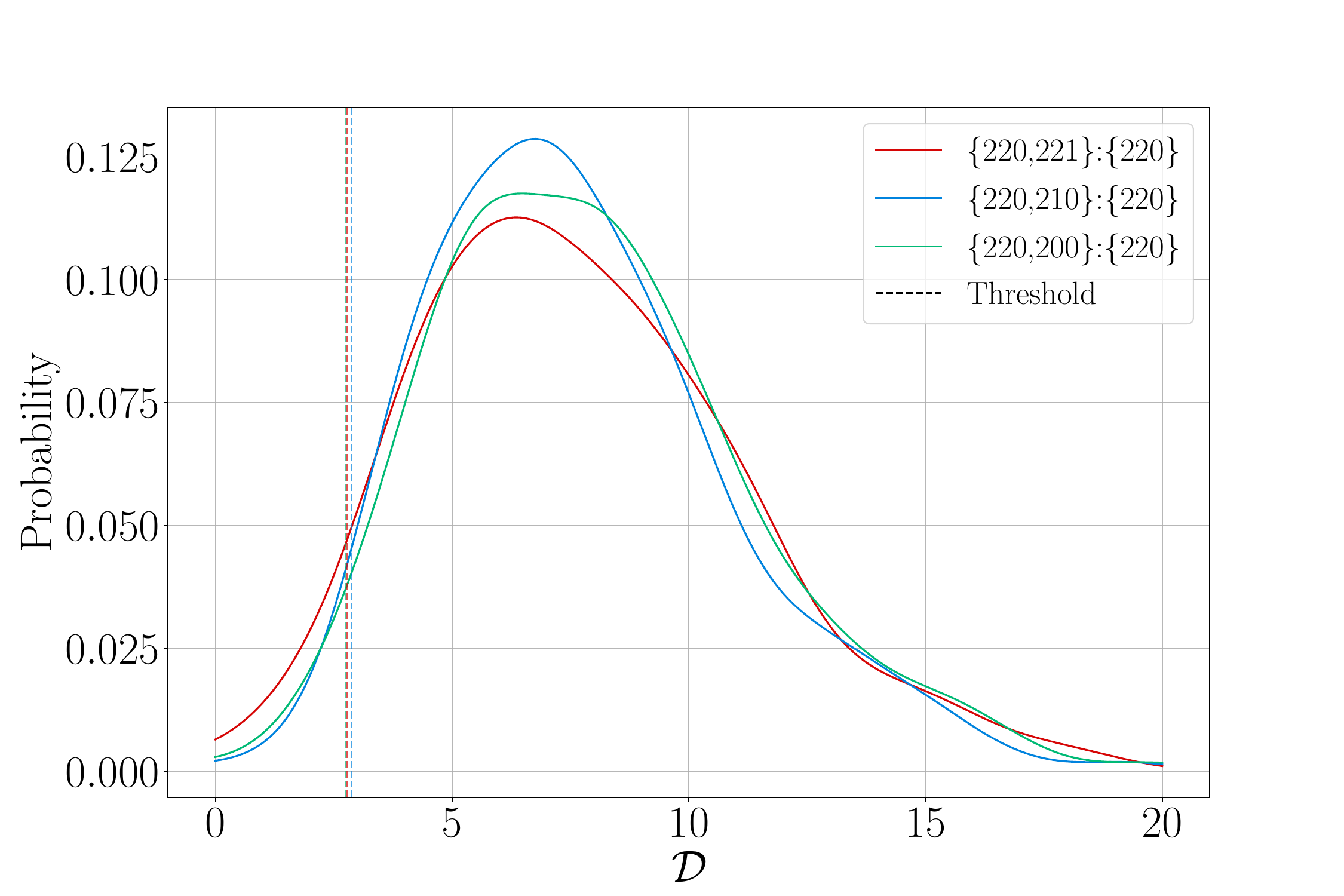}}
    \subfloat[\centering Distribution of $\mathcal{D}$ \label{fig:220+221_detStats}]{\includegraphics[width=0.5\textwidth]{BF_threshold/det_stat_diff_modes.pdf}}\hfill
    \subfloat[\centering Distribution of $p(M_{\rm inj}, \chi_{\rm inj})$ \label{fig:220+221_ppPlots}]{\includegraphics[width=0.5\textwidth]{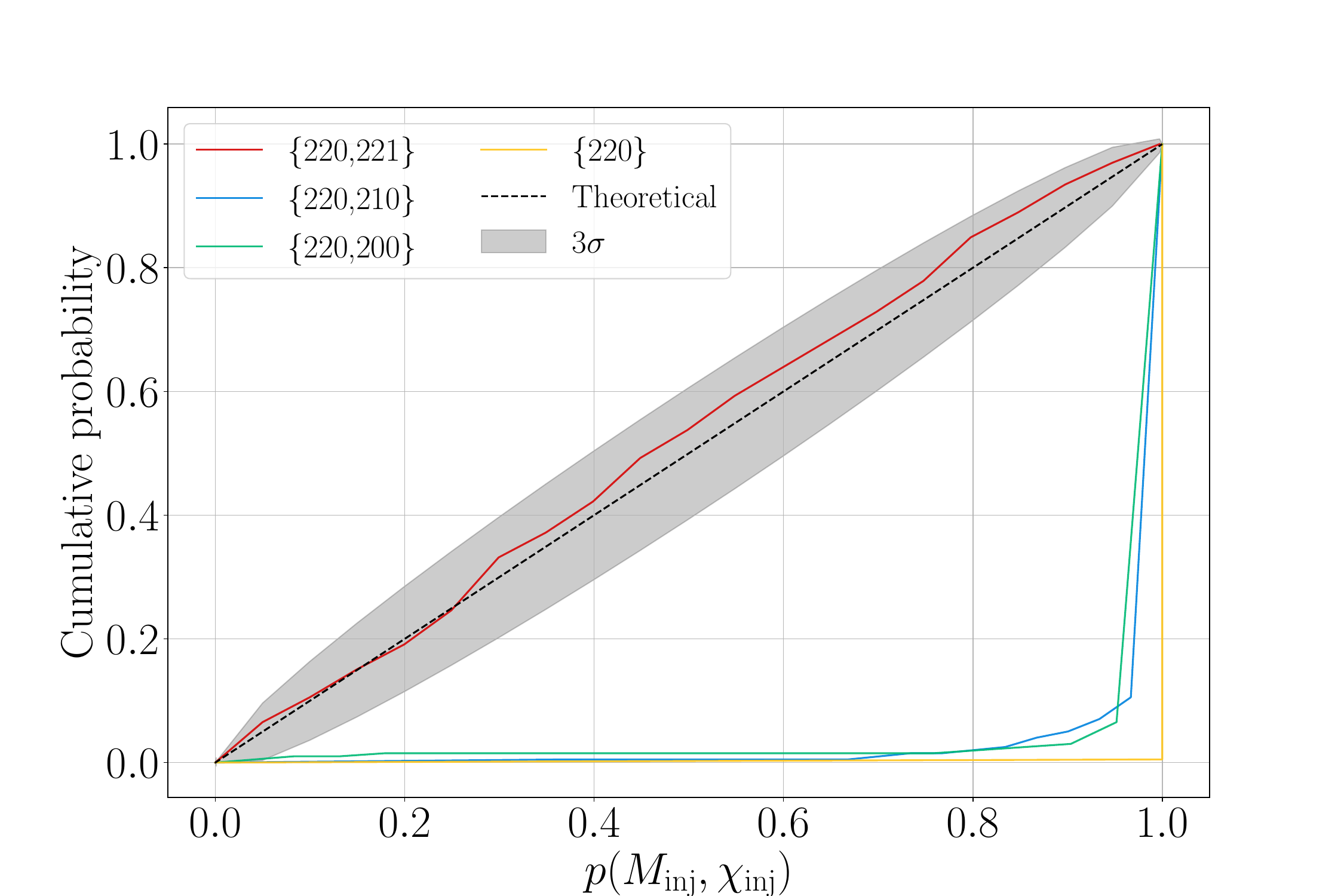}}\hfill
    \caption{Recovered $\mathcal{D}$ and $p(M_{\rm inj}, \chi_{\rm inj})$ from 200 injections of a \{220,221\} ringdown signal (see parameters in Table \ref{table:220+221_params}; with amplitudes scaled to yield SNR$_{221}=16$). Different colors correspond to different mode models (see legend). Panel (a) shows the probability distribution of $\mathcal{D}$ with different mode models. Panel (b) plots the cumulative probability distributions of $p(M_{\rm inj}, \chi_{\rm inj})$; the theoretical prediction of an accurate model with $3\sigma$ bounds is plotted for comparison. Each synthetic signal is injected into colored Gaussian noise of one aLIGO detector at the design sensitivity.}
    \label{fig:220+221_diffModes}
\end{figure}

The background study described in Sec.~\ref{sec:background} offers an approach to differentiate a true subdominant mode from fluctuations due to detector noise and, therefore, quantify the statistical significance of a potential subdominant mode. However, a subdominant mode present in the signal can also be falsely identified as a different QNM. To demonstrate this, we construct a \{220,221\} signal with parameters in Table \ref{table:220+221_params} and tune the amplitude so that the subdominant mode has SNR$_{221}=16$. We inject the signal into 200 noise realizations and compute $\mathcal{D}$ for both the correct mode model $\{220,221\}:\{220\}$ and two incorrect mode models, $\{220,210\}:\{220\}$ and $\{220,200\}:\{220\}$. Fig~\ref{fig:220+221_detStats} show that all three hypotheses yield similar distributions and all have $\mathcal{D}>\mathcal{D}^{\rm thr}$ in over 95\% realizations. The similarity between distributions occurs because all mode models are able to fit a majority of the SNR of the subdominant mode. We, therefore, cannot determine the correct mode on the basis of $\mathcal{D}$ alone at this SNR.

We use the posterior quantile of the injected parameters $p(M_{\rm inj}, \chi_{\rm inj})$ [Eq.~\eqref{eq:posterior_quantile}] as an additional criterion to identify the true subdominant mode when multiple hypotheses are above threshold. We expect the true mode hypothesis to result in lower posterior quantiles compared to other mode hypotheses. We verify this in Fig~\ref{fig:220+221_ppPlots} which shows the cumulative probability of $p(M_{\rm inj}, \chi_{\rm inj})$, i.e. the probability that $p(M_{\rm inj}, \chi_{\rm inj})$ is less than a given value on the x-axis: $P\left[p(M_{\rm inj}, \chi_{\rm inj})<x\right]$ where $P$ denotes probability. The plot demonstrates that the true \{220,221\} hypothesis is more likely to recover a lower $p(M_{\rm inj}, \chi_{\rm inj})$ than the other mode hypotheses which are biased to higher $p(M_{\rm inj}, \chi_{\rm inj})$. The plot also shows that the $\{220,221\}$ model hypothesis agrees with the theoretical expectation \cite{cook2006, talts2020}: 
\begin{gather}
    P\left[p(M_{\rm inj}, \chi_{\rm inj})<x\right] = x \, , \label{eq:cdf}
\end{gather}
within the $3 \sigma$ level. Eq.~\eqref{eq:cdf} holds for any signal containing a specific set of QNMs when it is recovered with the correct hypothesis. Incorrect models, however, are always biased to higher $p(M_{\rm inj}, \chi_{\rm inj})$. Using $p(M_{\rm inj}, \chi_{\rm inj})$ is, therefore, a robust way to identify the correct subdominant mode when there are multiple candidates. 

For real GW observations, we use the IMR-inferred remnant BH mass and spin $p(M_{\rm rem}, \chi_{\rm rem})$ to compute the posterior quantile. While this makes the analysis conditional on the IMR analysis, it is necessary to identify the correct subdominant mode at current ringdown SNRs [$\sim \mathcal{O}(10)$]. In a higher SNR regime, the distributions of $\mathcal{D}$ between the true model and an incorrect model are more distinct, which allows us to conduct IMR-agnostic analysis. We defer the investigation of the SNRs required for IMR-agnostic analysis to future work.  

%==========================================================================
\subsection{Workflow}
\label{sec:workflow}
\begin{figure*}[htb]
\includegraphics[width=\textwidth,clip=true]{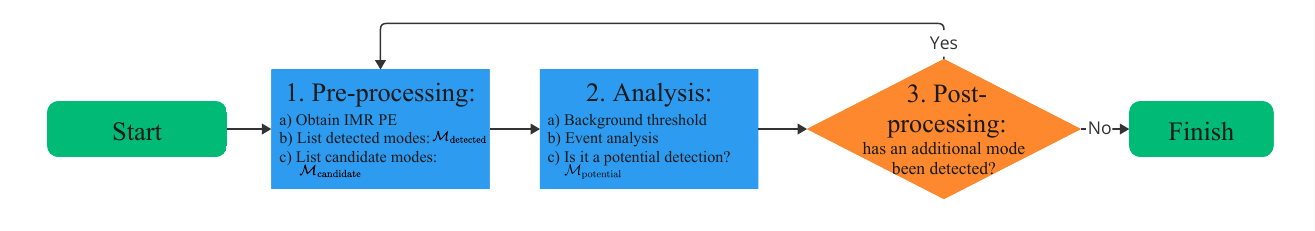}
  \caption{Flow chart describing the ringdown analysis steps for a GW event using the QNM rational filter.}
  \label{fig:QNMRF_flowchart}
\end{figure*}
We now formulate a complete workflow to analyze a GW event and identify all detectable ringdown modes using the rational filter. The workflow is outlined in Fig.~\ref{fig:QNMRF_flowchart}, and we describe it in detail below. 

\emph{Step 1. Pre-processing:} We first \emph{obtain the IMR PE} results, which are used as inputs by the rational filter. In particular, the sky position of the event (right ascension and declination) and geocentric start time are used to align the merger time at different detectors. The remnant BH mass and spin inferred by the IMR analysis are also needed in later steps when we compare the results from the rational filter analysis to those from the IMR. We then \emph{list detected modes: $\mathcal{M}_{\rm detected}$}. This list contains the ringdown modes identified in the data by the rational filter and initially consists only of the 220 mode (the dominant QNM for quasi-circular binaries and so assumed to be always detected, although see \cite{zhu2023} for precessing binaries), but is updated throughout the procedure. In principle, the analysis can simply start with an empty list and test the 220 mode as the first dominant mode. However, unless there are reasons to do this (e.g., analyzing a high-eccentricity merging system, obtaining unexpected results for initial analysis with the rational filter, etc.), we start by assuming that the 220 mode is the existing dominant mode. We now \emph{list candidate modes $\mathcal{M}_{\rm candidate}$} which could be the next most dominant mode. While the number of QNMs excited after merger can be arbitrarily large, in practice, only a small number of subdominant modes are potentially detectable with high confidence for current-generation detectors. We start with an exhaustive list of these modes, e.g., \{221, 210, 200, 330, 440\} and any others as needed for each individual event suggested by other ringdown pipelines or surrogate models \cite{pacilio2024}. The rational filter method is computationally efficient and thus allows us to test an exhaustive list of mode candidates. 

\emph{Step 2. Analysis:} We now iterate through the modes in $\mathcal{M}_{\rm candidate}$. For each candidate mode, we test if it is the next most dominant QNM by comparing two hypotheses; $\mathcal{H}$ where the modes in $\mathcal{M}_{\rm detected}$ exist and additionally the chosen candidate mode also exists, and $\mathcal{H}'$ where only $\mathcal{M}_{\rm detected}$ exist. As we iterate through $\mathcal{M}_{\rm candidate}$, the identity of the extra added mode changes, but we only ever test one additional mode at a time. Given the chosen candidate mode, we perform the $\mathcal{H}:\mathcal{H'}$ background study and compute the \emph{threshold} $\mathcal{D}^{\rm thr}$ as described in Sec~\ref{sec:FAP}. We then perform the \emph{event analysis} as described in Sec.~\ref{sec:background} and compute $\mathcal{D}_{\mathcal{H}:\mathcal{H'}}$. For both hypotheses, we also compute the joint posterior quantile on which the maximum a-posteriori value of the remnant BH mass and spin obtained from the IMR analysis $p(M_\textrm{\tiny IMR}, \chi_\textrm{\tiny IMR})$ lies. After these are computed, for each candidate mode being tested, we check \emph{is it a possible detection?} A candidate is classified as a potential detection if $\mathcal{D}>\mathcal{D}^{\rm thr}$ \emph{and} the posterior quantile decreases when including this candidate mode, i.e. adding the candidate mode makes the analysis more consistent with IMR \footnote{While other statistics aside from the posterior quantile may be a better indicator of the agreement between results from ringdown and IMR (e.g. how much of the 90\% credible regions obtained from the rational filter and the IMR analyses overlap), we defer a detailed study of different statistics to future work.}. If the candidate mode satisfies these criteria, it is added to a potential detections mode list $\mathcal{M}_{\rm potential}$. Otherwise, it is discarded. Regardless of whether the candidate is a potential detection or not, we repeat the procedure in Step 2 for each of the candidate modes in $\mathcal{M}_{\rm candidate}$ before continuing. 

\emph{Step 3. Post-processing:} If $\mathcal{M}_\textrm{potential}$ has at least one mode, an additional ringdown mode has been identified. In this case, we go back to Step 1 and add the top mode---the one yields the largest $\mathcal{D}$ in Step~2 and leads to the largest decrease in $p(M_\textrm{\small IMR}, \chi_\textrm{\small IMR})$---to $\mathcal{M}_\textrm{detected}$. We then look for the next subdominant mode and use the remaining (if any) modes in $\mathcal{M}_\textrm{potential}$ as the new candidate modes in $\mathcal{M}_\textrm{candidate}$. We additionally add the next overtone corresponding to the newly confirmed mode to $\mathcal{M}_\textrm{candidate}$ (i.e., if the 330 mode is confirmed, add the 331 mode to the candidate mode list). If, however, $\mathcal{M}_\textrm{potential}$ does not contain any modes, then no additional subdominant modes can be detected. The final confirmed ringdown modes list is then $\mathcal{M}_\textrm{detected}$, and we compute the total ringdown SNR for these modes (as described in Sec.~\ref{sec:SNR}). 

%The procedure described here cannot be independent of IMR analysis as the geocentric start time of the event is required. However, 
The procedure described above can alternatively be performed agnostic of the remnant mass and spin inferred by IMR, which means skipping the posterior quantile calculation in Step 2. In this case, the background study and threshold calculation can only be done using fully agnostic injections. However, to achieve better sensitivity, especially in the low-SNR regime, the full flow chart should be followed, and the thresholds should be computed using either IMR-conditional injections, restricted priors, or both, as described in Sec.~\ref{sec:improving_detectability}. 

%==========================================================================
\section{NR simulations}
\label{sec:NR}
We apply the procedure described in Sec.~\ref{sec:analysis_procedure} to analyze two NR waveforms from the Simulating eXtreme Spacetimes (SXS) catalog \cite{mroue2013, boyle2019}. We validate the method assuming the QNMs in the signal are not fully known a-priori. 
% \sma{Since we know the amplitudes of different modes that are injected, shall we provide an expected ranking of modes? This would be a strong cross-check for our algorithm. I guess the simplest way is to plot the envelope of each harmonic as a function of time, then we can visually see the ranking at every moment.}

\subsection{SXS:BBH:0305}
\label{sec:0305}
We first analyze the GW150914-like NR waveform, SXS:BBH:0305 \cite{lovelace2016}. We inject the $h_{2, \pm 2}$ harmonics into colored Gaussian noise at the aLIGO design sensitivity \cite{LVK_noise_curves} for two LIGO detectors (with inclination angle, $\iota$, and polarization angle, $\psi$, given by the maximum likelihood values inferred for GW150914 by the GWTC-2.1 PE \cite{theligoscientificcollaborationandthevirgocollaboration2024, collaboration2022}). We set the remnant mass in the detector frame to $M_{\rm inj}=68.5M_\odot$ and tune the amplitude such that the signal has a network ringdown SNR$^{\rm 2-ifo}=59.96$. The signal of 4 s is injected into noise with a sampling rate of $16384$~Hz. We select $\Delta t_0 \in [0,20]M_{\rm inj}$\footnote{We note that the precise onset of ringdown is ambiguous and remains a topic of ongoing study (e.g., \cite{thrane2017,bhagwat2018,giesler2019,baibhav2023,nee2023,giesler2024}). A detailed investigation of this issue is beyond the scope of this work.}, take 0.2 s of data starting at each $\Delta t_0$, and downsample it to 4096~Hz.

\begin{figure}[htb]
\includegraphics[width=\columnwidth,clip=true]{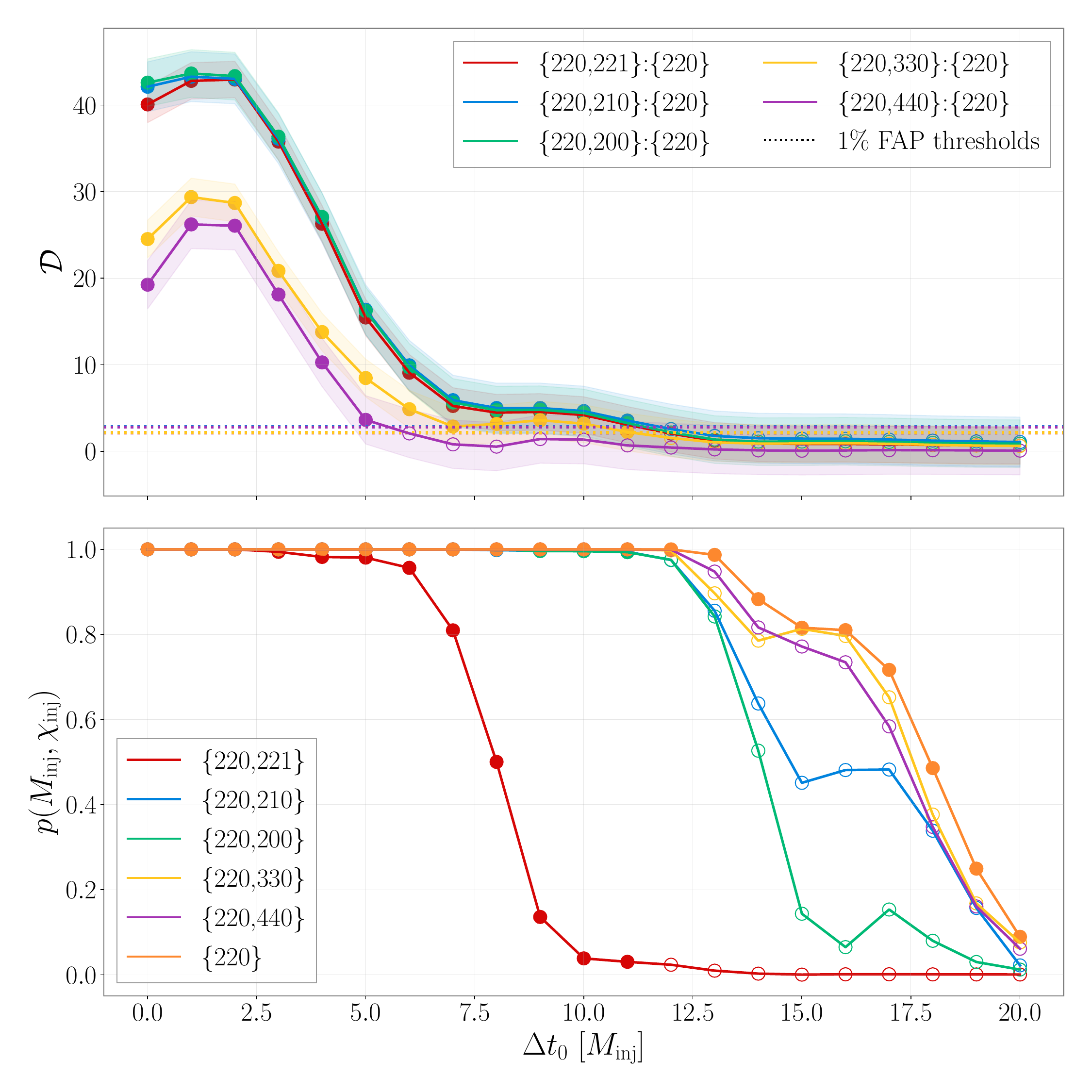}
  \caption{Resulting $\mathcal{D}$ and $p(M_{\rm inj}, \chi_{\rm inj})$ when testing different candidates of the second ringdown mode against the \{220\}-only hypothesis. The filled and unfilled markers correspond to $\mathcal{D} > \mathcal{D}^{\rm thr}$ and $\mathcal{D} < \mathcal{D}^{\rm thr}$, respectively. The shaded bands around $\mathcal{D}$ indicate the 99\% error bars due to
  potential noise fluctuations. The synthetic signal is constructed using the $h_{2, \pm 2}$ harmonic of the SXS:BBH:0305 waveform and injected into colored Gaussian noise with a network ringdown SNR of 59.86.}
  \label{fig:0305_2mode}
\end{figure}

Following Fig.~\ref{fig:QNMRF_flowchart}, in Step 1, we assume that the dominant QNM has already been detected and start with $\mathcal{M}_{\rm detected} = \{220\}$. The candidate mode list for the second ringdown mode is: $\mathcal{M}_{\rm candidate} = \{221, 210, 200, 330, 440\}$. In Step 2, we iterate through all candidates for the secondary mode. The resulting $\mathcal{D}$ and $p(M_{\rm inj}, \chi_{\rm inj})$ are shown in Fig.~\ref{fig:0305_2mode}. The dotted lines are the thresholds calculated as described in Sec.~\ref{sec:FAP}. The colored bands around $\mathcal{D}$ indicate the error bars due to potential noise fluctuations (i.e., corresponding to the 1-$\sigma$ range of the background $\mathcal{D}$ distribution); see Appendix~\ref{sec:BF_over_time} for more details. All of the candidate modes (independently analyzed) yield $\mathcal{D}>\mathcal{D}^{\rm thr}$ and also show a decrease of $p(M_{\rm inj}, \chi_{\rm inj})$ for $\Delta t_0 \leq 13M_\text{inj}$. Thus, they are accepted to the potential detection list, i.e., $\mathcal{M}_{\rm potential} = \{221, 210, 200, 330, 440\}$. In Step 3, we compare the different modes in $\mathcal{M}_{\rm potential}$ and find that the $\mathcal{D}$ for the secondary mode to be the 330 or 440 QNM are markedly lower than those for the 221, 210, and 200 modes. In the lower panel of Fig~\ref{fig:0305_2mode}, we also find that the posterior quantile of the \{220,221\} hypothesis is lower than the other modes for all times and leads to $p(M_{\rm inj}, \chi_{\rm inj}) \approx 0$ at $\Delta t_0 = 10 M$. Hypotheses with other candidate modes only have $p(M_{\rm inj}, \chi_{\rm inj}) \sim 0$ at much later times ($\Delta t_0 \sim 20 M$), when $\mathcal{D}$ values of the candidate modes have already fallen below the threshold and the posterior quantiles are no longer reliable. Therefore, the 221 QNM is the identified secondary mode. We return to Step 1 and update the detected and candidate mode lists to be $\mathcal{M}_{\rm detected} = \{220, 221\}$ and $\mathcal{M}_{\rm candidate} = \{222, 210, 200, 330, 440\}$. 

\begin{figure}[htb]
\includegraphics[width=\columnwidth,clip=true]{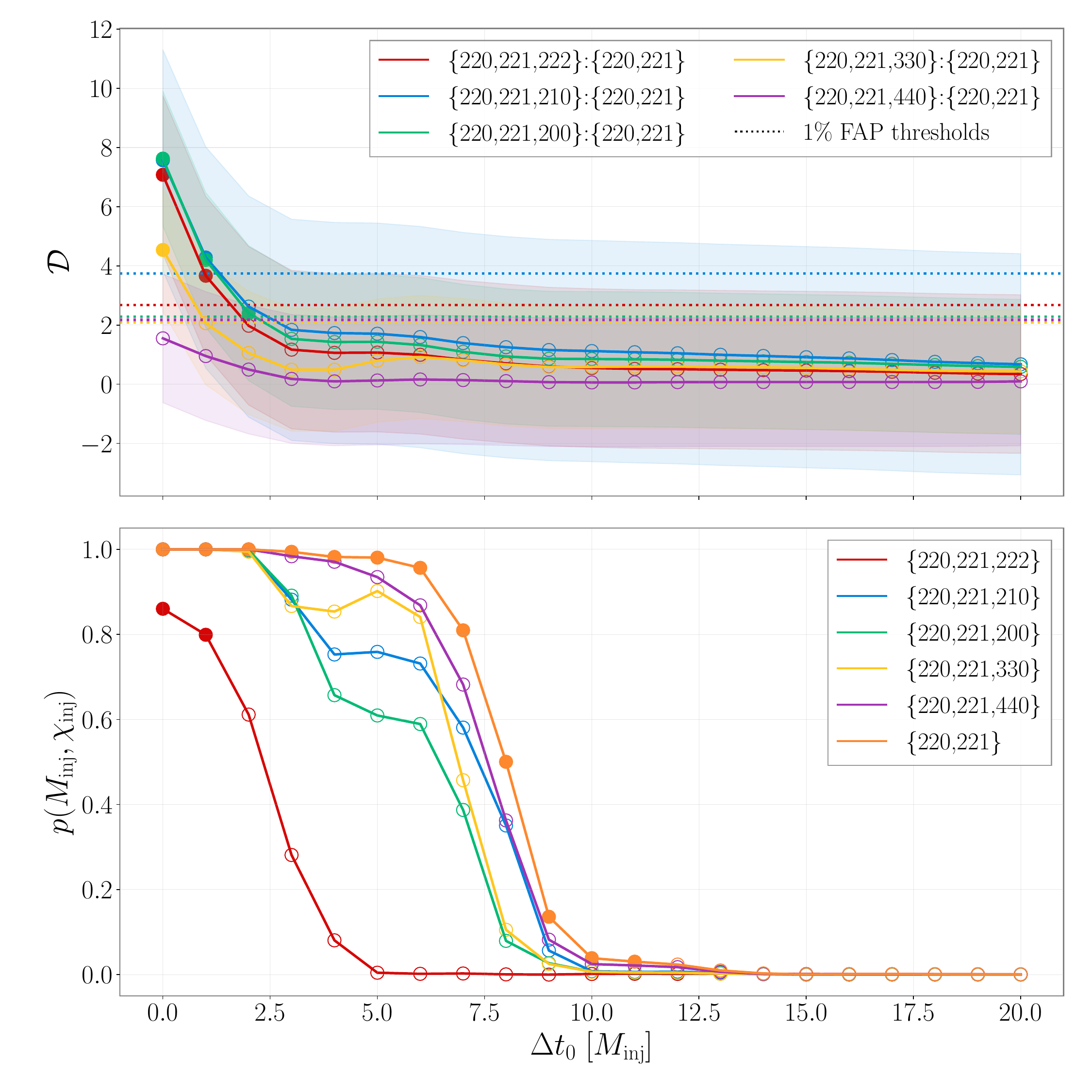}
  \caption{(Similar to Fig.~\ref{fig:0305_2mode}) Resulting $\mathcal{D}$ and $p(M_{\rm inj}, \chi_{\rm inj})$ when testing different candidates of the third ringdown mode compared to a \{220,221\} hypothesis.}
  \label{fig:0305_3mode}
\end{figure}

We now perform Step 2 for a third ringdown mode and show the results in Fig.~\ref{fig:0305_3mode}. In the upper panel, we see that the 440 candidate mode is not a potential detection as its $\mathcal{D}$ falls below the threshold at any start time.
%(i.e. the error bar intersects zero at all other times) so we discard it in Step 6. 
All other modes are added to $\mathcal{M}_{\rm potential}$. In the lower panel, we compare all potential modes following Step 3. We identify the third ringdown mode to be the 222 QNM because its $p(M_{\rm inj}, \chi_{\rm inj})$ are significantly lower than the others at all times, although the $\mathcal{D}$ values are comparable among all hypotheses. We now return to Step 1 and update the mode lists $\mathcal{M}_{\rm detected} = \{220,221,222\}$ and $\mathcal{M}_{\rm candidate} = \{223, 210, 200, 330\}$ to search for a fourth mode. However, no candidate yields $\mathcal{D} > \mathcal{D}^{\rm thr}$. Thus, we complete the analysis and estimate the network ringdown SNR, yielding 59.81. We compare the estimated SNR to the true injected SNR and find $\Delta \text{SNR} = 0.05$, demonstrating that we have recovered the majority of the signal power. 

\subsection{SXS:BBH:1107}
We now study a high-mass-ratio system (using the SXS:BBH:1107 waveform) with initial parameters $m_1/m_2 = 10$, $\chi_1 = \chi_2 = 10^{-7}$, where $m_1, m_2$ and $\chi_1, \chi_2$ are the initial masses and spins of the two progenitor BHs. We inject a synthetic signal using the remnant BH properties $M_{\rm inj} = 68.5M_\odot$, $\chi_{\rm inj} = 0.261$ with a ringdown SNR$^{\rm 2-ifo}=38.41$. In such a high-mass-ratio system, we expect that the higher-order angular modes are more strongly excited. We inject the $h_{2, \pm 2}$ and $h_{3, \pm 3}$ modes with ($\iota=\pi/4, \psi=0$) into colored Gaussian noise (two aLIGO detectors) as described in Sec.~\ref{sec:0305}. 
%As before, the signal is injected with a sampling rate of $16,384$ Hz and a duration of $4$ s with zeros padded to both ends of the NR waveform. 

\begin{figure}[htb]
\includegraphics[width=\columnwidth,clip=true]{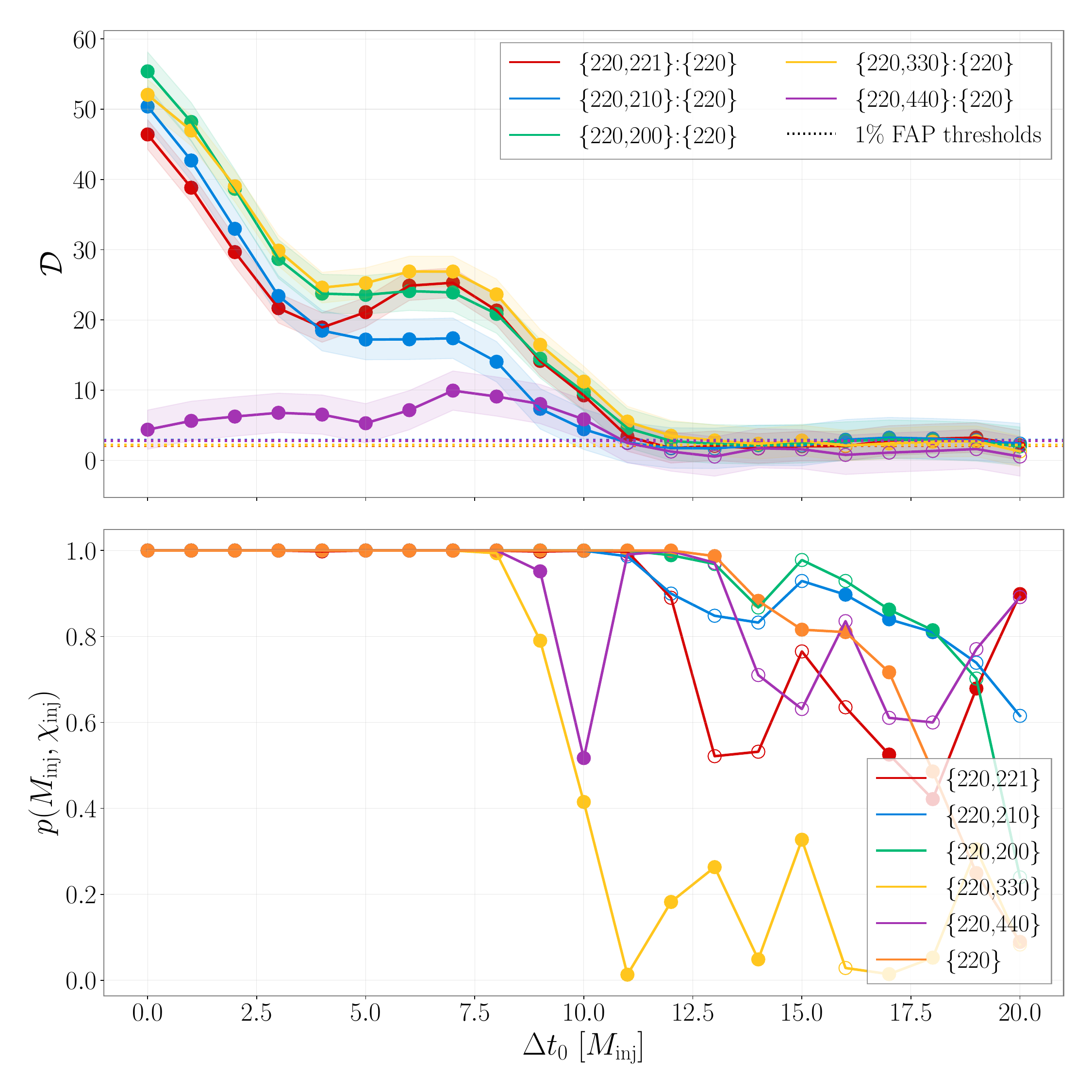}
  \caption{Resulting $\mathcal{D}$ and $p(M_{\rm inj}, \chi_{\rm inj})$ when testing different candidates of the second ringdown mode against the \{220\}-only hypothesis. The synthetic signal is constructed using the $h_{2, \pm 2}$ and $h_{3, \pm 3}$ harmonics of the SXS:BBH:1107 waveform and injected into colored Gaussian noise with a network ringdown SNR of 38.15.}
  \label{fig:1107_2mode}
\end{figure}

Again, we carry out the analysis following the steps in Fig.~\ref{fig:QNMRF_flowchart}. In Step 1, we start with $\mathcal{M}_{\rm detected} = \{220\}$ and $\mathcal{M}_{\rm candidate} = \{221, 210, 200, 330, 440\}$. The results of Step 2 are shown in Fig.~\ref{fig:1107_2mode}. The upper panel shows all candidate modes yield $\mathcal{D}>\mathcal{D}^{\rm thr}$. The lower panel shows all candidate modes also reduce $p(M_{\rm inj}, \chi_{\rm inj})$ compared to the $\mathcal{M}_{\rm detected}$, so they are all added to $\mathcal{M}_{\rm potential}$. In Step 3, we identify 330 mode as the top candidate for two reasons. First, $\mathcal{D}$ for 330 (yellow) is only slightly lower than that for 200 (green), and they are comparable within the uncertainty range. 
Second, adding the 330 QNM yields $p(M_{\rm inj}, \chi_{\rm inj}) \sim 0$ for $\Delta t_0 > 11 M_\textrm{inj}$, but adding the 200 QNM generally does not produce low $p(M_{\rm inj}, \chi_{\rm inj})$ values.
%\sma{Do we want to provide some 2D posterior plots? Maybe in appendix?} \ls{Agree providing contour examples in appendix would be good}
%that the BF for the secondary mode to be the 330 mode is second highest only to the 200 mode but that the posterior quantile of the 330 reaches and fluctuates around zero at $\Delta t_0 = 11M_f$ but does not do so for the 200 at any $\Delta t_0 < 20M_f$. 
We conclude that 330 mode is the secondary QNM and return to Step 1, updating the mode lists to be $\mathcal{M}_{\rm detected} = \{220, 330\}$ and $\mathcal{M}_{\rm candidate} = \{331, 221, 210, 200, 440\}$.  

\begin{figure}[htb]
\includegraphics[width=\columnwidth,clip=true]{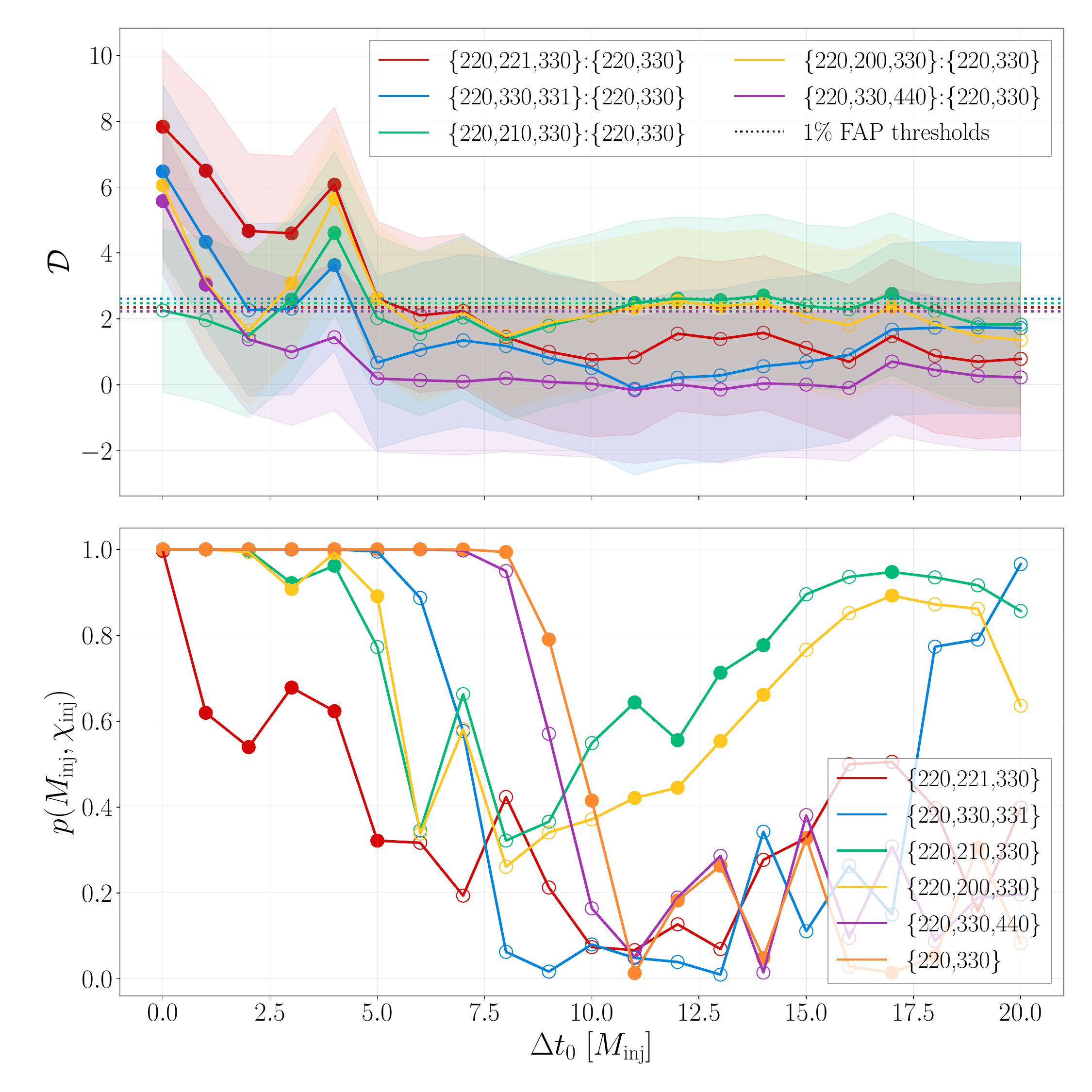}
  \caption{(Similar to Fig.~\ref{fig:1107_2mode}) Resulting $\mathcal{D}$ and $p(M_{\rm inj}, \chi_{\rm inj})$ when testing different candidates for the third ringdown mode compared to a \{220,330\} hypothesis.}
  \label{fig:1107_3mode}
\end{figure}

Fig.~\ref{fig:1107_3mode} shows the results from the next iteration to identify the third mode. Although all $\mathcal{D}$ values are comparable and above threshold for the majority of the time, the 221 mode yields the lowest $p(M_{\rm inj},\chi_{\rm inj})$ at times $\Delta t \le 7 M_{\rm inj}$. We update $\mathcal{M}_{\rm detected} = \{220, 330, 221\}$ and $\mathcal{M}_{\rm candidate} = \{331, 222, 210, 200, 440\}$. In the next iteration, none of the candidates satisfy $\mathcal{D} > \mathcal{D}^{\rm thr}$, so we conclude that all the detectable ringdown modes are identified. The total SNR estimated for $\{220, 330, 221\}$ modes is 38.15 with an error $\Delta \, \text{SNR}=-0.14$. 

\section{GW150914 analysis}
\label{sec:GW150914}
In this section, we analyze the real event GW150914 which was previously studied using the rational filter in Refs. \cite{ma2023, ma2023a}, where evidence for the 221 mode was found. We re-analyze the event by taking into account the noise background and quantifying the statistical significance of the overtone mode. Here, we follow Fig.~\ref{fig:QNMRF_flowchart} and repeat the procedure as demonstrated in the NR simulations. 
We use the PE results from the IMR analysis (Step 1), set the geocentric start time of the signal to be $t_0 = 1126259462.4083$, and use the inferred sky location (right ascension $\alpha = 1.95$~rad and declination $\delta = -1.27$~rad) to account for the signal travel time between the two aLIGO detectors. 
We start the analysis with $\mathcal{M}_{\rm detected} = \{220\}$ and $\mathcal{M}_{\rm candidate} = \{221\}$. We only consider the 221 mode as a candidate because GW150914 has been extensively studied, and no analysis finds evidence of any other mode (including using the rational filter). 

\begin{figure}[htb]
\includegraphics[width=\columnwidth,clip=true]{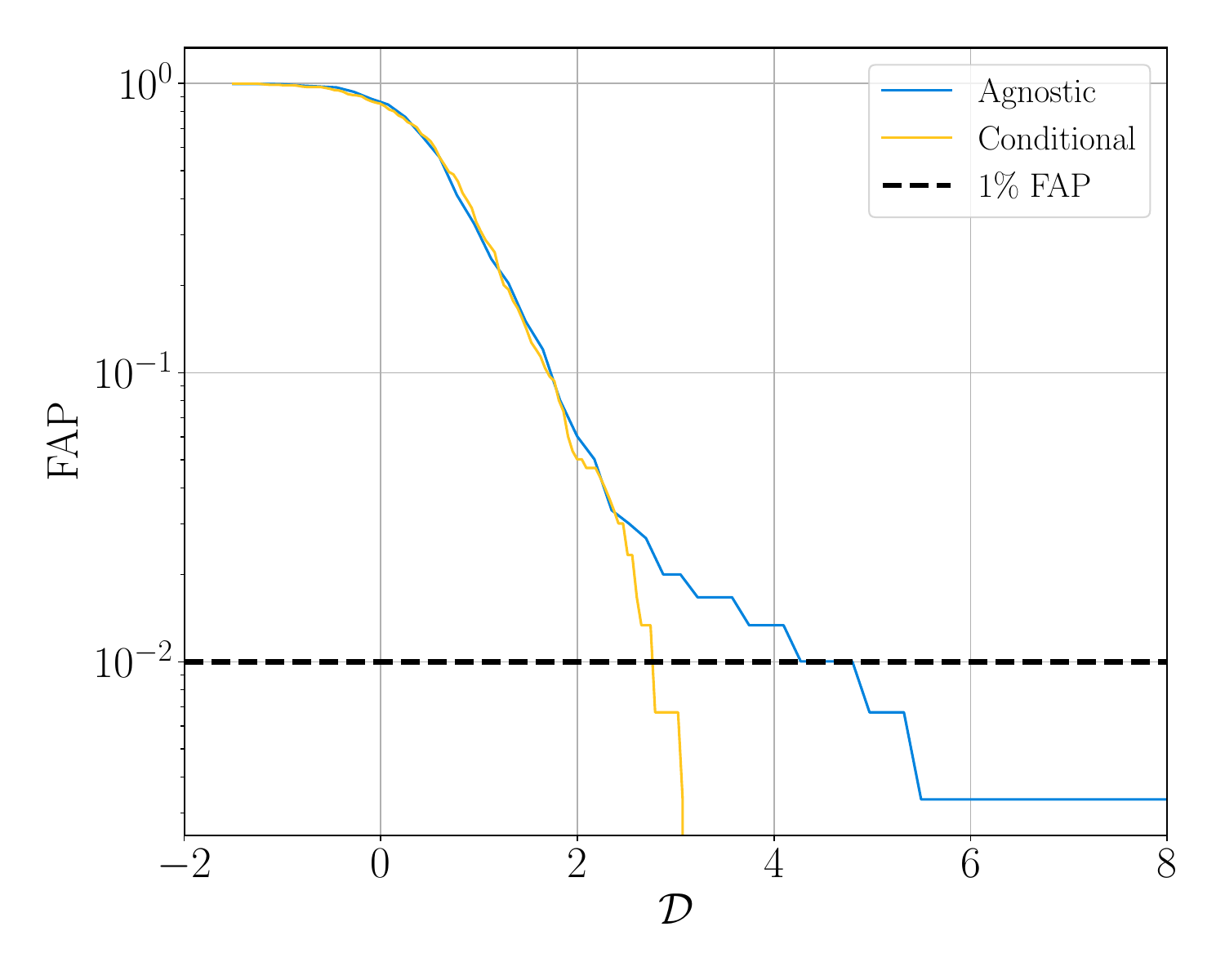}
  \caption{Background study of \{220,221\}:\{220\} for GW150914. Results are obtained by injecting a \{220\}-only waveform into 300 chunks of noise-only data around GW150914. As expected, the threshold corresponding to the 1\% FAP is lower with conditional injections than with agnostic injections.}
  \label{fig:GW150914_background}
\end{figure}

In Step 2, we calculate the $\mathcal{D}$ threshold when comparing the \{220,221\} and \{220\} hypotheses. Since the detector noise is non-stationary and non-Gaussian \cite{zackay2021, edy2021}, we take detector data in the 1.5 hours around the event time to evaluate the noise background. Specifically, we exclude 1 s of data around the event time and select 150 data chunks before and another 150 after the event, each spanning 16 s. For example, the first chunk after the event spans [$t_\text{start}$+0.5\,s, $t_\text{start}$+16.5\,s], and the next one spans  [$t_\text{start}$+16.5\,s, $t_\text{start}$+32.5\,s]. Similarly, we get all 300 noise data chunks. For each data chunk, we use the full 16 s to estimate the PSD and inject a \{220\} signal so that the peak strain of the injected signal occurs at the center of the data chunk, i.e., 8~s after the start. The $\mathcal{D}$ distribution of a \{220,221\} hypothesis compared to a \{220\} hypothesis is shown in Fig.~\ref{fig:GW150914_background}. The background is estimated both in an agnostic and a conditional way, as described in Sec.~\ref{sec:FAP}. The parameters for the conditional injections are sampled from the posterior distribution obtained in the IMR analysis~\cite{theligoscientificcollaborationandthevirgocollaboration2024, collaboration2022}. As expected, the threshold obtained is lower with the conditional injections than with the agnostic injections. 

We then analyze the GW150914 event itself, with a similar analysis configuration as that presented in Refs.~\cite{ma2023, ma2023a}. The analysis is performed using a 0.2~s long data stretch (i.e., the ``window width'' in \cite{ma2023, ma2023a}), and the data are conditioned in the same way (see Sec. II B of \cite{ma2023a}). However, this time we estimate the PSD by taking 16 s of data starting from 0.5 s after the peak of the event  (instead of 32 s around the event, including the event itself as described in \cite{ma2023, ma2023a}). We find this approach more robust in estimating PSDs where the ringdown signal has decayed sufficiently, but the noise behaviour remains similar to that during the event. In addition, we use a sampling rate of 4096~Hz instead of 2048 Hz used in Refs.~\cite{ma2023, ma2023a} to further reduce the impact of downsampling. 

\begin{figure}[htb]
\includegraphics[width=\columnwidth,clip=true]{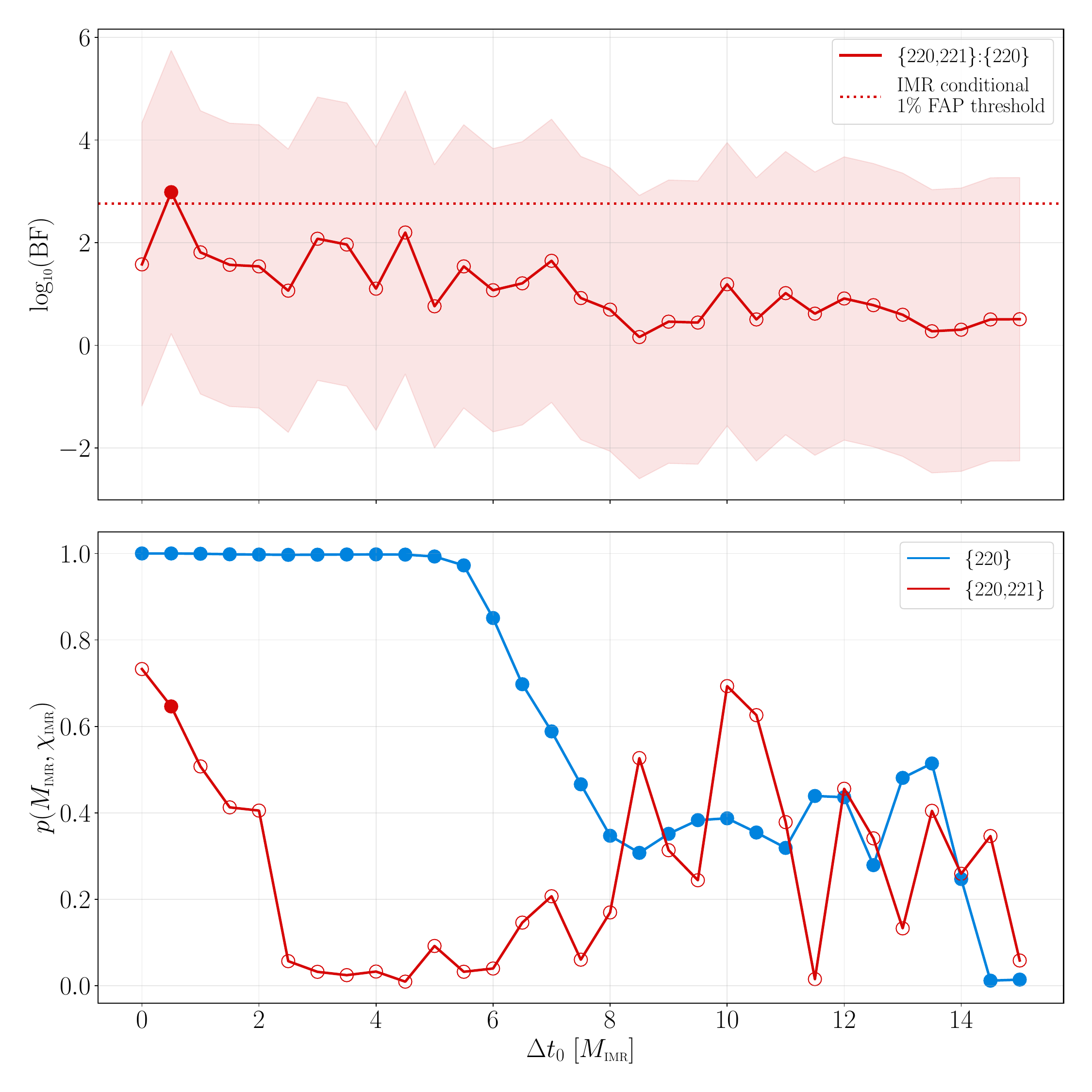}
  \caption{Resulting $\mathcal{D}$ and $p(M_{\rm IMR}, \chi_{\rm IMR})$ when comparing the \{220, 221\}:\{220\} hypotheses for the ringdown signal of GW150914. The dotted line is the threshold computed from IMR-conditional injections. The filled and unfilled markers correspond to $\mathcal{D} > \mathcal{D}^{\rm thr}$ and $\mathcal{D} < \mathcal{D}^{\rm thr}$, respectively. The shaded bands around $\mathcal{D}$ indicate the 99\% error bars due to potential noise fluctuations. }
  \label{fig:GW150914_analysis}
\end{figure}

Fig.~\ref{fig:GW150914_analysis} shows the results. The resulting $\mathcal{D}$ never exceeds the $\mathcal{D}^{\rm thr, \, agnostic}_{\{220,221\}:\{220\}}$ and is marginally above $\mathcal{D}^{\rm thr, \, conditional}_{\{220,221\}:\{220\}}$ only at $\Delta t_0 = 0.5 M_\textrm{IMR}$. For such a marginal detection, we do not test for the presence of a third ringdown mode. The estimated SNR for the \{220,221\} analysis at $\Delta t_0 = 0.5 M_f$ is 12.96. Thus, with our method, we conclude that there is only marginal evidence of a 221 overtone mode in GW150914 at an early time of the ringdown. These results are consistent with previous analysis of GW150914 with the rational filter which identified a $\mathcal{D}_{\{220,221\}:\{220\}}\approx2.8$ (presented in \cite{ma2023a} as a linear Bayes factor of $\approx 600$). 

%==========================================================================
\section{Conclusion}
\label{sec:conclusion}
In this work, we extend the QNM rational filter framework to improve the statistical identification of subdominant ringdown modes in GW signals. By employing a frequentist approach, we quantified the threshold based on a given FAP and proposed a systematic workflow for robustly identifying specific QNMs. The methodology is validated using synthetic signals built from NR waveforms, and we apply it to the real event GW150914 to assess the statistical significance of its first overtone.

The QNM rational filter provides an efficient tool for identifying ringdown modes without sampling over amplitudes and phases. The frequentist approach to background noise analysis enhances the robustness of mode identification by reducing false positives, thereby increasing confidence in subdominant mode detections. The improved workflow presented in this study paves the way for more robust tests of strong-field gravity and BH spectroscopy, offering new opportunities to verify the no-hair theorem and probe deviations from GR.

%==========================================================================
\begin{acknowledgments}
The authors would like to thank Katerina Chatziioannou, and Yanbei Chen for useful discussions at an early stage of this project, Max Isi, Christopher Moore, and Rahul Kashyap for helpful suggestions during an method review, and Harrison Siegel during a subsequent method review. This material is based upon work supported by NSF's LIGO Laboratory which is a major facility fully funded by the National Science Foundation. The authors are grateful for computational resources provided by the LIGO Laboratory and supported by National Science Foundation Grants PHY-0757058 and PHY-0823459. This research is supported by the Australian Research Council Centre of Excellence for Gravitational Wave Discovery (OzGrav), Project Numbers CE170100004 and CE230100016. LS is also supported by the Australian Research Council Discovery Early Career Researcher Award, Project Number DE240100206. Research at Perimeter Institute is supported in part by the Government of Canada through the Department of Innovation, Science and Economic Development and by the Province of Ontario through the Ministry of Colleges and Universities. EF acknowledges support from the Department of Energy under award number DE-SC0023101.
\end{acknowledgments}  
%==========================================================================
\appendix

\section{The rational filter as a hybrid Bayesian-like analysis}
\label{sec:semi-Bayesian}
Previous work on the rational filter described it as a Bayesian analysis \cite{ma2022, ma2023, ma2023a}. Here we provide clarification about the extent to which the rational filter is Bayesian. 

The likelihood in Eq.~\eqref{eq:likelihood} depends on the parameters of the rational filter: $M_f$ and $\chi_f$. The rational filter analysis and the contours computed in the $M_f$ and $\chi_f$ plane are Bayesian in this respect. However, when comparing the rational filter analysis to the usual time-domain analysis, it is shown in the supplementary material of Ref.~\cite{ma2023} that, for a single detector, the rational filter is closely connected to the usual time-domain likelihood when using the maximum likelihood estimation for mode amplitudes under the assumption of white noise. The maximum-likelihood estimation for amplitudes is a frequentist method\footnote{It is further shown in the supplementary material of Ref.~\cite{ma2023} that the maximum likelihood estimation for mode amplitudes is equivalent to marginalizing over the amplitudes when using an infinitely large flat prior---although this is formally an improper prior.} which is then combined with the Bayesian inference on $M_f$ and $\chi_f$. For multiple detectors, the rational filter is additionally incoherent in mode amplitudes across detectors (as discussed in Sec.~\ref{sec:1_mode}). 
%We thus refer to the rational filter as ``semi-Bayesian'' to distinguish it from the usual time-domain analysis. 
The detection statistic $\mathcal{D}$ defined in this work [Eq.~\eqref{eq:BF}] is thus directly analogous to a log Bayes factor but is distinct from the Bayes factor of time-domain analyses. It is still, however, capable of being used to distinguish between different mode hypotheses (as verified throughout the paper and specifically in Sec. \ref{sec:NR}). 
We thus refer to the rational filter as a hybrid Bayesian-like approach to distinguish it from the usual time-domain analysis.

Within this paper, we provide a method to quantify the statistical significance of a given $\mathcal{D}$ in order to distinguish true signal observations from noise more robustly. This improves on more generalized criteria for the necessary Bayes factors required for a statistical significant detection, such as those suggested in \cite{kass1995, jeffreys1998}. Specifically, the procedure outlined in Sec.~\ref{sec:FAP} better accounts for the detector noise behavior around a GW event and the impact from any non-Gaussian noise features. However, this work does not comment on the known issues of using Bayes factors when well-motivated priors are not known \cite{gelman2017, robert2007}. This work also does not comment on the time at which constant amplitude QNM fittings are valid, as discussed in \cite{giesler2024, baibhav2023}. 

% \section{Checking segment length and sampling rate}
% \label{sec:segment_length_and_srate}
% \begin{figure}[htb]
% \includegraphics[width=\columnwidth,clip=true]{BF_threshold/BFthresh_simulation_number.pdf}
%   \caption{The computed $\mathcal{D}^{\rm thr}$ for the \{220,221\}:\{220\} background analysis as a function of simulation number from 50 to 1000 injections for 3 different shuffles of the same set of injections. The estimated $\mathcal{D}^{\rm thr}$ is well approximated by the estimated using 200 injections and $\mathcal{D}^{\rm thr}$ does not change when increasing the injection number beyond 300. }
%   \label{fig:BF_threshold_convergence}
% \end{figure}

\section{Convergence of $\mathcal{D}^{\rm thr}$}
\label{sec:convergence_BF_thresh}
\begin{figure}[htb]
\includegraphics[width=\columnwidth,clip=true]{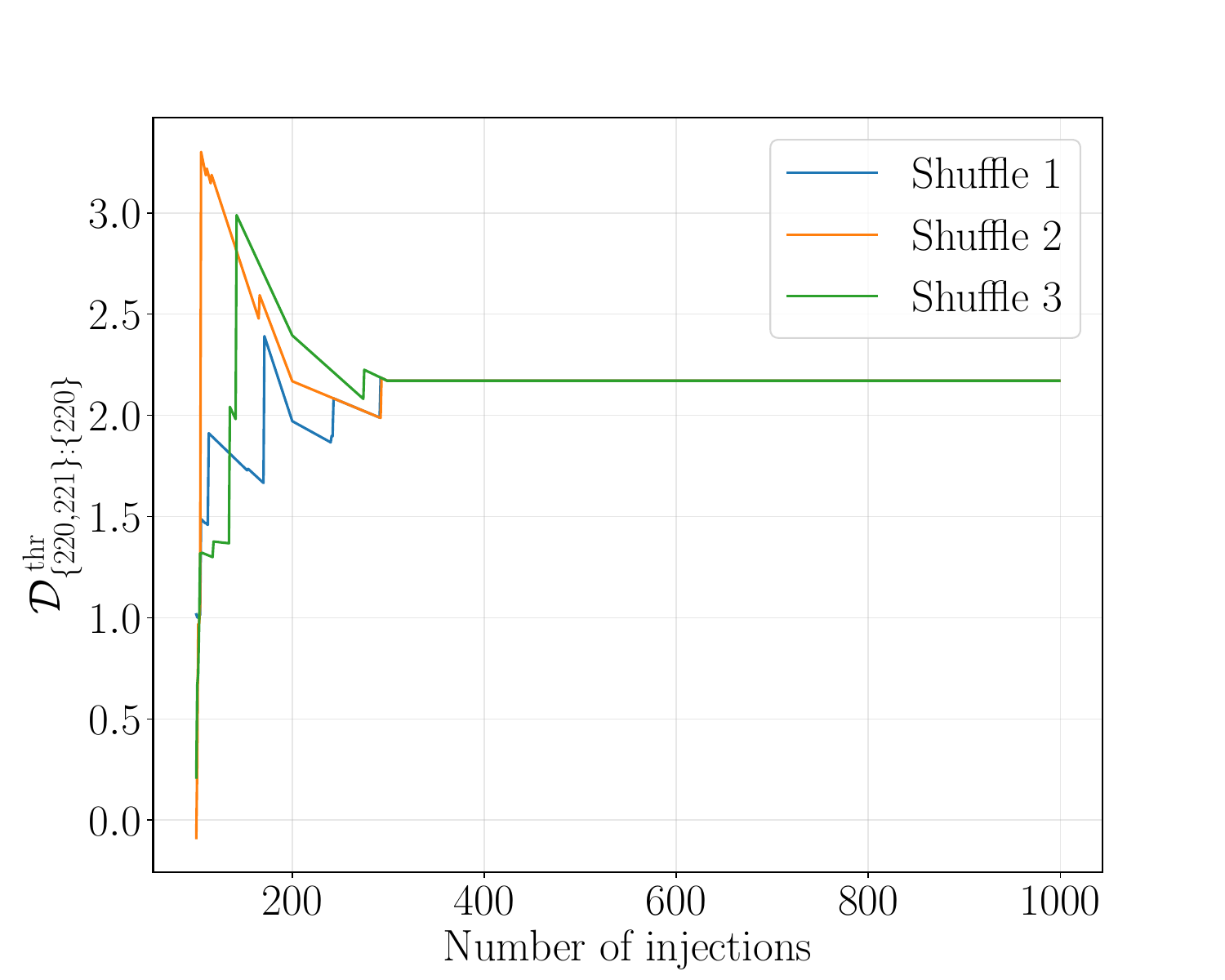}
  \caption{$\mathcal{D}^{\rm thr}$ for the \{220,221\}:\{220\} background study estimated using different numbers of injections (from 50 to 1000) for three different shuffles . }
  \label{fig:BF_threshold_convergence}
\end{figure}

To determine how many simulations are required in the background study to obtain a convergent $\mathcal{D}^{\rm thr}$, we carry out the \{220,221\}:\{220\} background analysis using 1000 injections. We consider how the threshold $\mathcal{D}^{\rm thr, \, 2-ifo}_{\{220+221\}:\{220\}}$ varies as we consider subsets of the injections of increasing size, starting from 50 simulations up to the full set of 1000. We also shuffle the order of the injections three times to check the robustness of the calculated threshold. Fig.~\ref{fig:BF_threshold_convergence} shows that 200 injections are able to obtain a fairly robust threshold, which remains constant when using $>300$ injections. 

The noise around real GW events is not guaranteed to be stationary or Gaussian which may affect the number of simulations required for a convergent estimate of $\mathcal{D}^{\rm thr}$. For real GW events, therefore, we would check convergence using the Gelman-Rubin statistic \cite{gelman1992} or other statistical tests (e.g., those described in \cite{PyMC}). 

\section{Error bars of $\mathcal{D}$}
\label{sec:BF_over_time}
Here, we seek to quantify the error bar of a calculated $\mathcal{D}$. Recall that $\mathcal{D}$ is the ratio between the evidence for two different mode hypotheses [Eq.~\eqref{eq:BF}]. It is common to approximate the evidence of a given hypothesis by its maximum likelihood value \cite{cornish2011} (here we neglect the Occam's factor). Let us now quantify the variance of the maximum likelihood for a given hypothesis using Eq.~\eqref{eq:likelihood} and Eq.~\eqref{eq:inner_prod}:
\begin{gather}
    \sigma^2 = {\rm Var}\Bigl( -\frac{1}{2}\Bigl \langle (s + n)^F \Bigl|(s + n)^F \Bigr\rangle \Bigr).
\end{gather}
For simplicity, we factor out the $-\frac{1}{2}$ term, use the linearity of the filter and the linearity of the inner product [defined in Eq.~\eqref{eq:inner_prod}], and get:
\begin{align}
    4 \sigma^2 &= {\rm Var}\Bigl( \bigl \langle s^F \bigl|s^F \bigr\rangle + 
    2 \bigl \langle s^F \bigl|n^F \bigr\rangle + 
    \bigl \langle n^F \bigl|n^F \bigr\rangle\Bigr). \\
\intertext{Using the identity $\rm Var(A+B) = \rm Var(A) + \rm Var(B) + 2\,\rm Cov(A, B)$ (where Cov denotes the covariance operator) we get:}
    \begin{split}
        4 \sigma^2 = \, &{\rm Var}\Bigl( \bigl \langle s^F \bigl|s^F \bigr \rangle \Bigr ) 
        + {\rm Var}\Bigl( 2 \, \bigl \langle s^F \bigl|n^F \bigr \rangle + \bigl \langle n^F \bigl|n^F \bigr \rangle \Bigr ) \\
        & + 2 \, {\rm Cov} \Bigl (\bigl \langle s^F \bigl|s^F \bigr \rangle , 2 \, \bigl \langle s^F \bigl|n^F \bigr\rangle + 
    \bigl \langle n^F \bigl|n^F \bigr\rangle \Bigr ).
    \end{split}
\end{align}
The first term ${\rm Var}\Bigl( \bigl \langle s^F \bigl|s^F \bigr \rangle \Bigr )=0$ because $s^F$ is a constant (not a distribution). Similarly, the covariance term is zero because its first argument $\langle s^F \bigl|s^F \bigr \rangle$ is a constant. We are, therefore, left with:
\begin{align}
    4 \sigma^2 = & \,{\rm Var}\Bigl( 2\, \bigl \langle s^F \bigl|n^F \bigr \rangle + \bigl \langle n^F \bigl|n^F \bigr \rangle \Bigr ) \\
    \begin{split}
        = &\; 4 \; {\rm Var} \Bigl( \langle s^F \bigl| n^F \rangle \Bigr ) 
        + {\rm Var} \Bigl( \langle n^F \bigl| n^F \rangle \Bigr )\\
        &+ 2 \; {\rm Cov} \Bigl( \langle s^F \bigl| n^F \rangle, \langle n^F \bigl| n^F \rangle \Bigr).
    \end{split}
\end{align}
Here we can use the Cauchy-Schwarz inequality for random variables to bound the covariance term:
\begin{align}
    \begin{split}
        4 \sigma^2 \leq & \; 4 \; {\rm Var} \Bigl( \langle s^F \bigl| n^F \rangle \Bigr ) 
        + {\rm Var} \Bigl( \langle n^F \bigl| n^F \rangle \Bigr )\\
        &+ 2 \; \sqrt{{\rm Var} \Bigl( \langle s^F \bigl| n^F \rangle \Bigr) {\rm Var} \Bigl( \langle n^F \bigl| n^F \rangle \Bigr)}.
    \end{split} \label{eq:sigma_expression}
\end{align}
Let us consider the first term and rewrite it in terms of the expectation $E$:
\begin{align}
    {\rm Var} \Bigl( \langle s^F \bigl| n^F \rangle \Bigr ) = E \Bigl[\langle s^F \bigl| n^F \rangle^2 \Bigr] - E \Bigl[\langle s^F \bigl| n^F \rangle  \Bigr]^2.
\end{align}
The second term is zero because $\langle s^F \bigl| n^F \rangle$ is the inner product between a constant ($s^F$) and a random distribution with zero mean $n^F$ so it's expectation value is zero. We re-express the first term using the commutativity of the inner product for real-valued inputs:
\begin{align}
    {\rm Var} \Bigl( \langle s^F \bigl| n^F \rangle \Bigr ) &= E \Bigl[\langle s^F \bigl| n^F \rangle \langle n^F \bigl| s^F \rangle \Bigr] \notag \\
    &= \langle s^F \bigl| s^F \rangle \, , \label{eq:var_sf_nf}
\end{align}
where we use an identity from Eq.~(3) of \cite{vallisneri2008}. Eq.~\eqref{eq:var_sf_nf} is exactly the residual optimal SNR of the misfiltered signal, denoted by SNR$_\text{res}$. This is related to the fitting factor using an alternative definition \cite{cornish2011}:
\begin{gather} 
\text{FF}^2 = \text{max} \left(1 - \frac{\text{SNR}_\text{res}^2}{\text{SNR}^2} \right) , \label{eq:FF_alt} 
\end{gather}
where the maximum is taken over the prior parameter space and corresponds to the maximum likelihood values used to approximate the evidence. We are interested in finding the FF between the two hypotheses for calculating $\mathcal{D}$. While we cannot compute the FF analytically, as explained in more detail in Appendix~\ref{sec:2nd_mode_SNR}, we can numerically verify that the FFs between the hypotheses we have tested are all within the range of [0.98, 1) for a variety of systems. Since the ringdown SNR of current GW observations is $\mathcal{O}$(10), the residual SNR of the misfiltered waveforms are $\ll 1$. When substitute this into Eq.~\eqref{eq:sigma_expression} it eliminates the first and third terms and that remains is:
\begin{gather}
    \sigma^2 \leq \; {\rm Var} \Bigl( -\frac{1}{2}\langle n^F \bigl| n^F \rangle \Bigr ).
\end{gather}
This is exactly the quantity computed in Sec.~\ref{sec:FAP} to determine the FAP. Therefore, the 99\% error bar around the computed $\mathcal{D}$ is exactly $\mathcal{D}^{\rm thr}$. 

\section{FF between non-trivial hypotheses}
\label{sec:2nd_mode_SNR}

Here, we provide a detailed explanation about why the rational filter cannot compute the SNR of individual ringdown modes. We first review the effect of applying a rational filter built to remove the ($\ell'$, $m'$, $n'$) mode to the $(\ell, m, n)$ QNM [$h(t)$ in Eq.~\eqref{eq:QNM_def}]. The filtered time-series becomes \cite{ma2023a}:
\begin{eqnarray}
    h^F(t) &=& A_{\ell mn}B_{\ell mn}^{\ell'm'n'} e^{-(t - t_0)/\tau_{\ell mn}} \nonumber \\
    &&\times \cos \left[ 2\pi f_{\ell mn} (t-t_0) + \phi_{\ell mn} + \varphi^{\ell'm'n'}_{\ell mn} \right],
\label{eq:misfiltered}
\end{eqnarray}
where the amplitude $B^{\ell mn}_{\ell'm'n'}$ and phase $\varphi_{\ell'm'n'}^{\ell mn}$ terms can be computed from the filter: 
\begin{equation}
    B^{\ell mn}_{\ell'm'n'} e^{i\varphi^{\ell mn}_{\ell'm'n'}} = \mathcal{F}_{\ell'm'n'}(\omega_{\ell mn}) \, . \label{eq:B_varphi_def}
\end{equation}
So the application of an incorrect filter reduces the amplitude of the signal by a factor of $B^{\ell mn}_{\ell'm'n'}$ and introduces a phase shift of $\varphi^{\ell mn}_{\ell'm'n'}$. 

\begin{figure}[htb]
\settowidth{\imagewidth}{\includegraphics{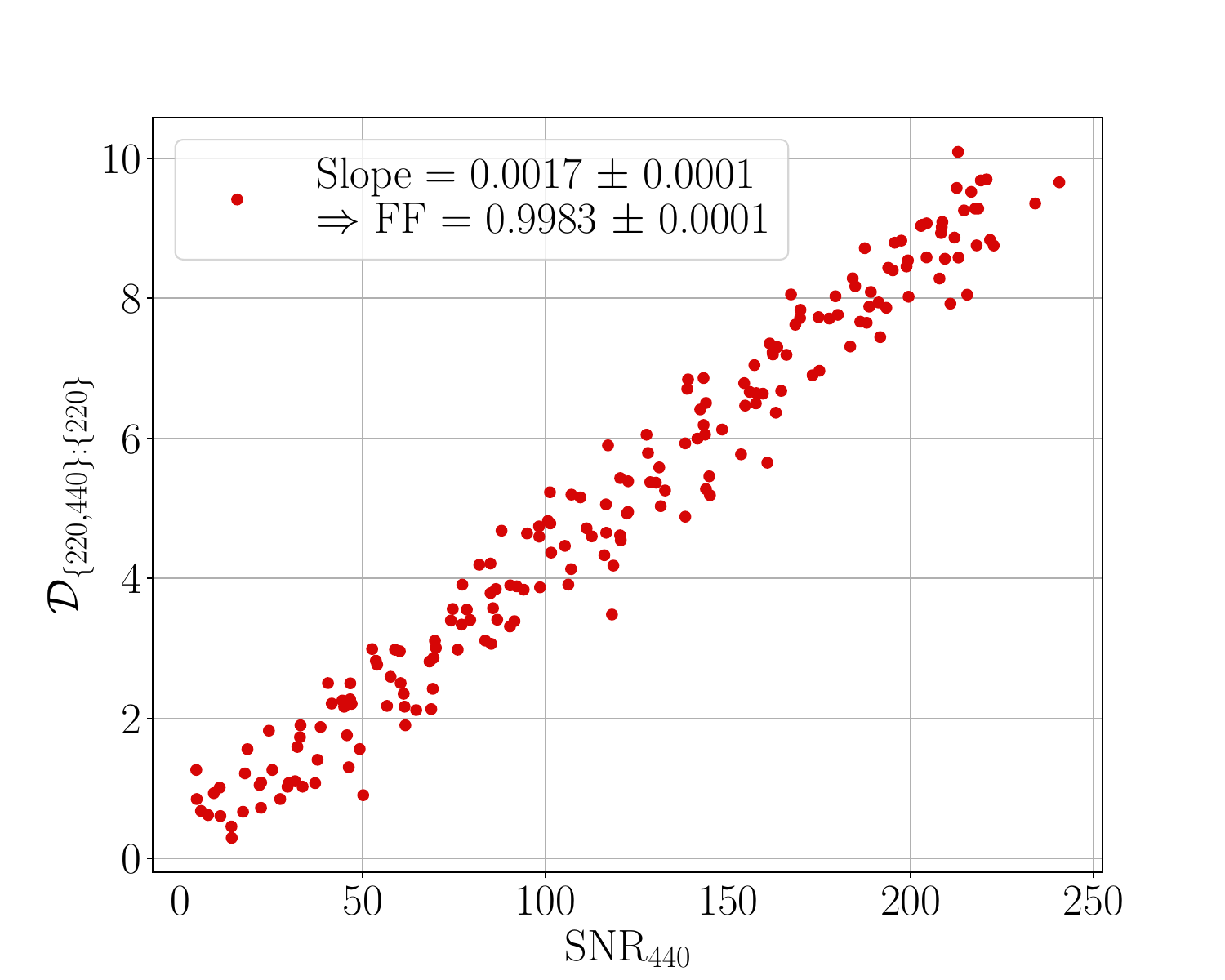}}
    \subfloat[\centering Fixed injection parameters.\label{fig:FF!=0_fixed}]{\includegraphics[width=0.5\textwidth]{SNR_scaling/SNR_scaling-220+440_220_constFF.pdf}}\hfill
    \subfloat[\centering Randomized injection parameters. \label{fig:FF!=0_randomized}]{\includegraphics[width=0.5\textwidth]{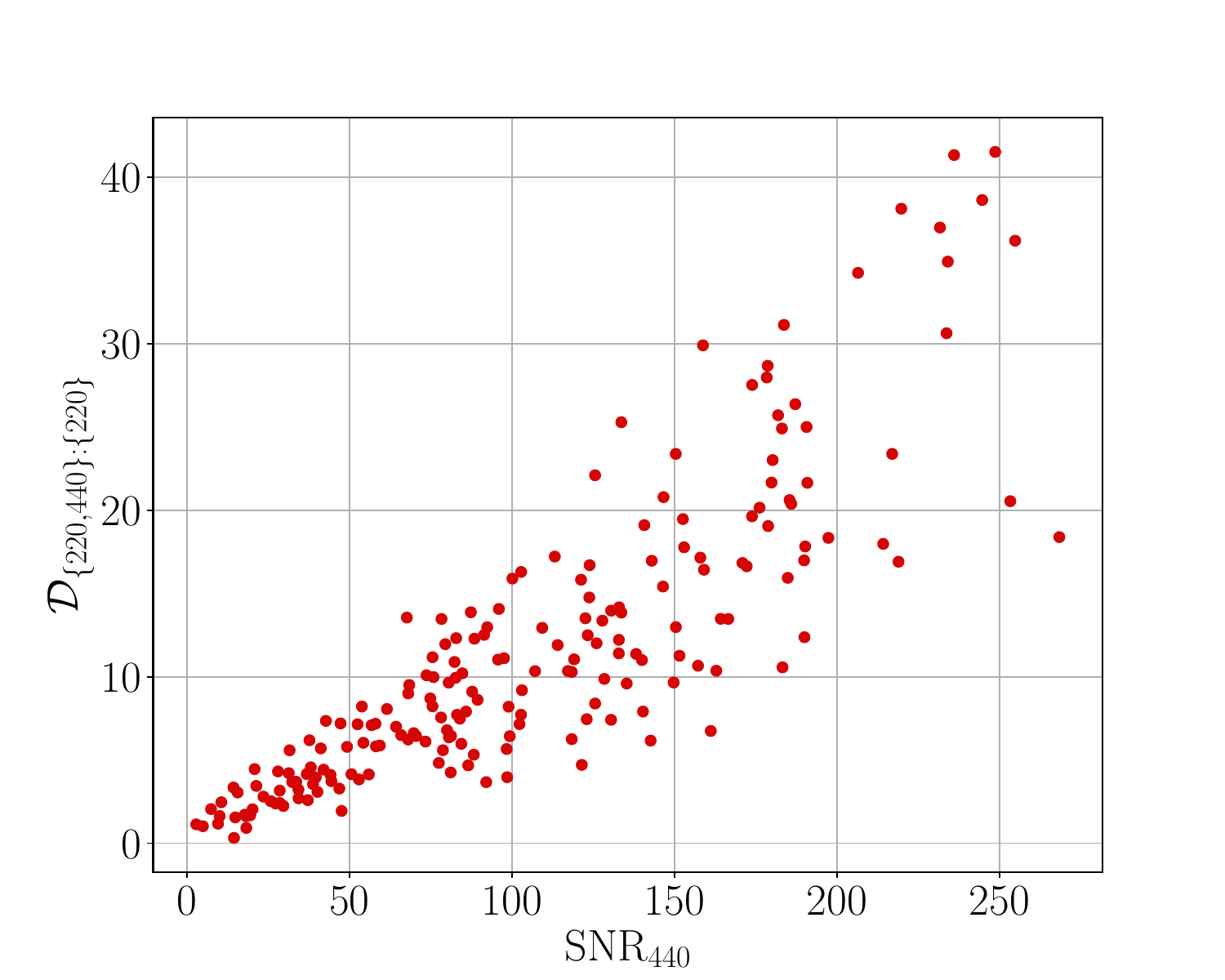}}\hfill
    \caption{$\mathcal{D}_{\{220,440\}:\{220\}}$ as a function of the 440-mode SNR. }
    \label{fig:FF!=0}
\end{figure}

Let us now consider an injection of \{220,440\} QNMs and analyze it under two hypotheses: a \{220,440\} mode hypothesis and a \{220\} hypothesis. For simplicity, we inject a weak 440 mode so that there is little systematic bias in the inferred remnant mass and spin when using the incorrect \{220\} model, but relax this condition later. The \{220\} filter with the maximum likelihood completely removes the 220 mode but also partially filters out the 440 mode. From Eq.~\eqref{eq:BF_vs_SNR}, we could compute the SNR of the 440 mode if we could determine the FF between the \{220,440\}:\{220\} models. We can try to do so by considering the alternative definition of the FF in Eq.~\eqref{eq:FF_alt} \cite{cornish2011}.
The maximum value corresponds to the filter that produces the maximum likelihood in the prior parameter space. In the specific scenario considered here, SNR$_\text{res}$ is the residual 440 mode SNR after it is partially filtered out by the maximum likelihood \{220\} filter. Since the rational filter does not affect the noise power or spectrum \cite{ma2023}, we can consider the energy of the waveforms instead of the SNR:
\begin{gather}
\frac{\text{SNR}_\text{res}^2}{\text{SNR}^2} = \frac{\int_0^\infty |h_t^F|^2 dt}{\int_0^\infty |h(t)|^2 dt} \, , \label{eq:energy_ratio}
\end{gather}
where recall that $h^F_t$ is the misfiltered waveform in the time domain and is computed analogously using Eq.~\eqref{eq:filtered_data}. In the case of a $(\ell, m, n)$ mode being incorrectly filtered by the rational filter of the mode $(\ell', m', n')$, we can analytically compute Eq.~\eqref{eq:energy_ratio} using Eq.~\eqref{eq:misfiltered} to get:
\begin{widetext}
\begin{equation}
\frac{\text{SNR}_\text{res}^2}{\text{SNR}^2} 
= \frac{\left(B_{440}^{220}\right)^ 2(1+\omega_{440}^2\tau^2_{440} + \cos[2(\phi_{440}+\varphi_{440}^{220})]-\omega_{440} \tau_{440} \sin[2(\phi_{440}+\varphi_{440}^{220})])}
{\left(\sum_{\ell mn=\{220,440\}}1+\omega_{\ell mn}^2\tau_{\ell mn}^2 + \cos(2\phi_{\ell mn}) - \omega_{\ell mn}\tau_{\ell mn}\sin(2\varphi_{\ell mn}^{\ell'm'n'})\right)}\,.
\label{eq:FF_expression}
\end{equation}
\end{widetext}
Eq.~\eqref{eq:FF_expression} depends on $\phi_{440}$, and $M_{\rm inj}$ and $\chi_{\rm inj}$ of the system that control $B_{\ell mn}^{\ell'm'n'}$ and $\varphi_{\ell mn}^{\ell'm'n'}$ through Eq.~\eqref{eq:B_varphi_def}. No information about the phase $\phi_{440}$ is inferred by the rational filter, and the constraints on $M_{\rm inj}$ and $\chi_{\rm inj}$, for most realistic GW observations, are imprecise. If we relax our condition that the 440 mode is weak and hence does not bias the maximum likelihood parameter estimate from the \{220\} filter, then Eq.~\eqref{eq:FF_expression} becomes even more complex with a summation over the \{220,440\} modes in the denominator. Therefore, we cannot determine the values of Eq.~\eqref{eq:FF_expression} and thus the FF between any non-trivial hypotheses. This means that even when comparing the same hypotheses, the FF remain constant only when $\phi_{440}$, $M_{\rm inj}$, and $\chi_{\rm inj}$ are fixed, as illustrated in Fig.~\ref{fig:FF!=0}. Fig.~\ref{fig:FF!=0_fixed} shows that if the injection parameters are fixed, there is a linear relationship between SNR and $\sqrt{\mathcal{D}/\ln (10)}$---the FF between the hypotheses is a well-defined value. In Fig.~\ref{fig:FF!=0_randomized}, however, when the injection parameters are randomized, there is a large uncertainty in the slope (i.e. the FF between hypotheses varies) so that there is no clear mapping relationship between $\sqrt{\mathcal{D}/\ln(10)}$ and the single mode SNR. 

%\clearpage

%%%%%%%%%%%%%%%%%%%%%%%%%%%%%%%%%%%%%%%%%%%%%%%%%%%%%%%%%%%%%%%%%%%%%%%%%%%%%%%
\def\bibsection{\section*{References}}
%%%%%%%%%%%%%%%%%%%%%%%%%%%%%%%%%%%%%%%%%%%%%%%%%%%%%%%%%%%%%%%%%%%%%%%%%%%%%%%
%\bibliographystyle{apsrev4-2-author-truncate.bst}
\bibliography{References.bib}

\end{document}